\documentclass[acmsmall,screen,nonacm,review=false,timestamp=false]{acmart}
\newif\ifcamready
\camreadytrue

\newif\ifdraft
\draftfalse

\usepackage{xcolor}
\usepackage{booktabs}
\usepackage{verbatim}
\usepackage{url}
\usepackage{graphicx}
\usepackage{enumitem}

\usepackage{hyperref}
\usepackage{titlesec}
\usepackage{caption}
\usepackage{tabularx}
\usepackage{amsmath}
\usepackage{pifont}
\usepackage{relsize}
\usepackage{units}
\usepackage{censor}
\usepackage{tablefootnote}
\usepackage{amsthm}
\usepackage[tight-spacing=true]{siunitx}

\usepackage{natbib}
\usepackage[font=small]{caption}

\usepackage[T1]{fontenc}

\usepackage{scalefnt}

\usepackage{listings}

\definecolor{linkcol}{rgb}{0,0,0}
\definecolor{citecol}{rgb}{0,0,0}
\definecolor{urlcol}{rgb}{0,0,0.75}

\hypersetup{%
	hidelinks,%
	breaklinks=true,%
	bookmarksnumbered=true,%
	hypertexnames=true,%
}

\newenvironment{prettylist}{
	\begin{list}{
		\footnotesize\raisebox{0pt}{\small\ding{121}}
	}{
		\setlength\topsep{2pt plus 1pt minus 1pt}
		\setlength\leftmargin{2em}
		\setlength\rightmargin{0pt}
		\setlength\itemsep{1pt plus.1pt}
		\setlength\parskip{0pt}
		\setlength\parsep{0pt}
		\setlength\itemindent{0pt}
	}
}{
	\end{list}
}

\newcommand\colname[1]{\textit{#1}}

\ifdraft
\definecolor{notecolor}{rgb}{0.8,0,0} 
\newcommand{\note}[1]{{\textcolor{notecolor}{[\textit{#1}]}}}
\else
\newcommand{\note}[1]{}
\fi

\titlespacing*{\section}{0pt}{1.25ex plus.5ex minus.2ex}{.75ex plus.25ex}
\titlespacing*{\subsection}{0pt}{1.25ex plus.5ex minus.2ex}{.75ex plus.25ex}
\titlespacing*{\subsubsection}{0pt}{.75ex plus.25ex minus.25ex}{.25ex plus.2ex}

\renewcommand\textless{\raisebox{0.5pt}{\relsize{-0.25}{<}}}

\newcommand\maxray{Edact-Ray}

\newcommand\strname{Str}
\newcommand\acrnname{Acrn}

\newcommand\wordname{Word}
\newcommand\fnname{FN}
\newcommand\lnname{LN}
\newcommand\filnname{FILN}
\newcommand\fixlnname{FI$\times$LN}
\newcommand\fnlnname{FNLN}
\newcommand\fnxlnname{FN$\times$LN}
\newcommand\ctryname{Ctry}
\newcommand\rgnname{Rgn}
\newcommand\natlname{Natl}

\newcommand\tfontname{Times}
\newcommand\afontname{Arial}
\newcommand\bfontname{Calibri}

\newcommand\nadjshortname{Un}
\newcommand\wxiishortname{W07}
\newcommand\wxxishortname{W19}

\newcommand\uniname{Uniformly distributed}

\newcommand\dnsaNumDocs{678}

\newcommand\recapNumDocs{\num{2.5e4}}
\newcommand\oigNumDocs{\num{1.9e4}}

\newcommand\dnsaPg{\num{1.2e4}}
\newcommand\foiaPg{\num{4.9e5}}
\newcommand\oigPg{\num{5.2e5}}
\newcommand\recapPg{\num{3.4e5}}

\newcommand\dnsaNumDocsWRedact{7}

\newcommand\recapNumDocsWRedact{67}

\newcommand\numRECAPpages{$\approx10^{8}$}

\newcommand\dnsaMetaRedact{235}
\newcommand\foiaMetaRedact{\num{4.5e4}}
\newcommand\oigMetaRedact{\num{1.3e4}}

\newcommand\recapMetaRedact{1,221}

\newcommand\unadjDNSA{7}

\newcommand\unadjRECAP{7}

\newcommand\ocrDNSA{0}

\newcommand\ocrRECAP{224}

\newcommand\mswDNSAExact{14}

\newcommand\mswRECAPExact{119}

\newcommand\mswDNSANear{3}

\newcommand\mswRECAPNear{33}

\newcommand\unrecDNSA{214}

\newcommand\unrecFOIA{\num{1.3e4}}

\newcommand\unrecRECAP{838}

\newcommand\unrecOIG{\num{1e4}}

\newcommand\numOIGUnmatched{39}

\newcommand\numRECAPNameredactions{6,541}
\newcommand\numRECAPredactionDocs{710}

\newcommand\numRECAPnadj{327}

\newcommand\numRECAPocr{445}
\newcommand\numRECAPmsw{58}
\newcommand\numRECAPunk{5,691}
\newcommand\numRECAPNearMSW{20}

\newcommand\nameDictSizeComb{7,066,800}

\newcommand\numFOIAUnmatched{382}
\newcommand\numFOIAUniq{3}
\newcommand\numOIGUniq{0}
\newcommand\numFOIAavg{2,435}
\newcommand\numFOIAmed{494}
\newcommand\numOIGavg{4,260}
\newcommand\numOIGmed{1,081}

\newcommand\neltword{63,054} 
\newcommand\hxword{15.9} 
\newcommand\neltfn{100,364} 
\newcommand\hxfn{16.6} 
\newcommand\neltln{151,671} 
\newcommand\hxln{17.2} 
\newcommand\neltctry{566} 
\newcommand\hxctry{9.1} 
\newcommand\neltrgn{2,794,808} 
\newcommand\hxrgn{21.4} 
\newcommand\neltnatl{509} 
\newcommand\hxnatl{9.0} 
\newcommand\neltfiln{3,943,446} 
\newcommand\hxfiln{21.9} 
\newcommand\neltfnln{15,222,308,244} 
\newcommand\hxfnln{33.8} 
\newcommand\neltstr{90,706,185,230,812,058,042,928}
\newcommand\hxstr{76.3}
\newcommand\neltacrn{12,356,604}
\newcommand\hxacrn{23.6}

\renewcommand\neltacrn{\num{1.2e7}}
\renewcommand\neltstr{\num{9e22}}
\renewcommand\neltfiln{\num{3.9e6}}
\renewcommand\neltfnln{\num{1.5e10}}
\renewcommand\neltrgn{\num{2.8e6}}

\newcommand\noccnytword{432,896,070} 
\newcommand\hxnytword{12.3} 
\newcommand\noccnytfn{71,031,188} 
\newcommand\noccnytln{97,551,697} 
\newcommand\noccnytfiln{2,139,713} 
\newcommand\noccnytfnln{14,440,238} 
\newcommand\noccnytctry{3,905,371} 
\newcommand\hxnytctry{5.9} 
\newcommand\noccnytrgn{33,636,150} 
\newcommand\hxnytrgn{10.6} 
\newcommand\noccnytnatl{4,081,019} 
\newcommand\hxnytnatl{5.4} 
\newcommand\noccnytstr{791,209,093} 
\newcommand\hxnytstr{11.8} 
\newcommand\noccnytacrn{4,674,379} 
\newcommand\hxnytacrn{10.0} 

\newcommand\hyunytlnnadjbx{12.7} 
\newcommand\pgupctnytlnnadjbx{6\%} 
\newcommand\hyunytlnnadjax{8.3} 
\newcommand\pgupctnytlnnadjax{\textless1\%} 
\newcommand\hyunytlnnadjrx{2.9} 
\newcommand\hyunytlnnadjtx{8.2} 
\newcommand\pgupctnytlnnadjtx{\textless1\%} 
\newcommand\hyunytwordnadjbx{12.8} 
\newcommand\pgupctnytwordnadjbx{16\%} 
\newcommand\hywnytwordnadjbx{11.2} 
\newcommand\pgwpctnytwordnadjbx{74\%} 
\newcommand\hyunytwordnadjax{8.7} 
\newcommand\pgupctnytwordnadjax{2\%} 
\newcommand\hywnytwordnadjax{7.9} 
\newcommand\pgwpctnytwordnadjax{27\%} 
\newcommand\hyunytwordnadjrx{3.3} 
\newcommand\hywnytwordnadjrx{3.2} 
\newcommand\hyunytwordnadjtx{8.7} 
\newcommand\pgupctnytwordnadjtx{2\%} 
\newcommand\hywnytwordnadjtx{8.1} 
\newcommand\pgwpctnytwordnadjtx{29\%} 
\newcommand\hyunytfnnadjbx{12.3} 
\newcommand\pgupctnytfnnadjbx{8\%} 
\newcommand\hyunytfnnadjax{7.9} 
\newcommand\pgupctnytfnnadjax{\textless1\%} 
\newcommand\hyunytfnnadjrx{2.6} 
\newcommand\hyunytfnnadjtx{7.8} 
\newcommand\pgupctnytfnnadjtx{\textless1\%} 
\newcommand\hyunytstrnadjbx{12.1}

\newcommand\pgupctnytstrnadjbx{\textless1\%}

\newcommand\hyunytacrnnadjbx{11.4}

\newcommand\pgupctnytacrnnadjbx{\textless1\%}

\newcommand\hywnytstrnadjbx{10.5} 
\newcommand\pgwpctnytstrnadjbx{74\%} 
\newcommand\hywnytacrnnadjbx{9.1} 
\newcommand\pgwpctnytacrnnadjbx{81\%} 
\newcommand\hyunytstrnadjax{8.8}

\newcommand\pgupctnytstrnadjax{\textless1\%}

\newcommand\hyunytacrnnadjax{6.4}

\newcommand\pgupctnytacrnnadjax{\textless1\%}

\newcommand\hywnytstrnadjax{7.5} 
\newcommand\pgwpctnytstrnadjax{35\%} 
\newcommand\hywnytacrnnadjax{6.4} 
\newcommand\pgwpctnytacrnnadjax{43\%} 
\newcommand\hyunytstrnadjrx{0.2}

\newcommand\hyunytacrnnadjrx{0.2}

\newcommand\hywnytstrnadjrx{3.0} 
\newcommand\hywnytacrnnadjrx{2.0} 
\newcommand\hyunytstrnadjtx{8.2}

\newcommand\pgupctnytstrnadjtx{\textless1\%}

\newcommand\hyunytacrnnadjtx{6.5}

\newcommand\pgupctnytacrnnadjtx{\textless1\%}

\newcommand\hywnytstrnadjtx{7.6} 
\newcommand\pgwpctnytstrnadjtx{37\%} 
\newcommand\hywnytacrnnadjtx{6.5} 
\newcommand\pgwpctnytacrnnadjtx{44\%} 

\newcommand\hyuavgnytwordwxiitx{12.6}
\newcommand\hywavgnytwordwxiitx{10.9}

\newcommand\pguavgpctnytwordwxiitx{22\%}
\newcommand\pgwavgpctnytwordwxiitx{69\%}

\newcommand\hyuavgnytwordwxiiax{12.5}
\newcommand\hywavgnytwordwxiiax{10.6}

\newcommand\pguavgpctnytwordwxiiax{23\%}
\newcommand\pgwavgpctnytwordwxiiax{64\%}

\newcommand\hyuavgnytwordwxiibx{14.3}
\newcommand\hywavgnytwordwxiibx{11.8}

\newcommand\pguavgpctnytwordwxiibx{48\%}
\newcommand\pgwavgpctnytwordwxiibx{88\%}

\newcommand\hyuavgnytwordwxxitx{12.3}
\newcommand\hywavgnytwordwxxitx{10.6}

\newcommand\pguavgpctnytwordwxxitx{19\%}
\newcommand\pgwavgpctnytwordwxxitx{63\%}

\newcommand\hyuavgnytwordwxxiax{11.8}
\newcommand\hywavgnytwordwxxiax{10.2}

\newcommand\pguavgpctnytwordwxxiax{16\%}
\newcommand\pgwavgpctnytwordwxxiax{57\%}

\newcommand\hyuavgnytwordwxxibx{14.3}
\newcommand\hywavgnytwordwxxibx{11.7}

\newcommand\pguavgpctnytwordwxxibx{47\%}
\newcommand\pgwavgpctnytwordwxxibx{87\%}

\newcommand\hyuavgnytlnwxiitx{12.4}

\newcommand\pguavgpctnytlnwxiitx{11\%}

\newcommand\hyuavgnytlnwxiiax{12.2}

\newcommand\pguavgpctnytlnwxiiax{12\%}

\newcommand\hyuavgnytlnwxiibx{14.8}

\newcommand\pguavgpctnytlnwxiibx{33\%}

\newcommand\hyuavgnytlnwxxitx{11.7}

\newcommand\pguavgpctnytlnwxxitx{8\%}

\newcommand\hyuavgnytlnwxxiax{11.4}

\newcommand\pguavgpctnytlnwxxiax{7\%}

\newcommand\hyuavgnytlnwxxibx{14.6}

\newcommand\puwavgpctnytlnwxxibx{14\%}
\newcommand\pguavgpctnytlnwxxibx{30\%}

\newcommand\hyuavgnytfixlnnadjrx{2.9}

\newcommand\hyuavgnytfixlnnadjtx{8.5}

\newcommand\pguavgpctnytfixlnnadjtx{\textless1\%}

\newcommand\hyuavgnytfixlnnadjax{8.6}

\newcommand\pguavgpctnytfixlnnadjax{\textless1\%}

\newcommand\hyuavgnytfixlnnadjbx{12.8}

\newcommand\pguavgpctnytfixlnnadjbx{\textless1\%}

\newcommand\hyuavgnytfixlnwxiitx{13.1}

\newcommand\pguavgpctnytfixlnwxiitx{1\%}

\newcommand\hyuavgnytfixlnwxiiax{13.2}

\newcommand\pguavgpctnytfixlnwxiiax{2\%}

\newcommand\hyuavgnytfixlnwxiibx{15.4}

\newcommand\pguavgpctnytfixlnwxiibx{5\%}

\newcommand\hyuavgnytfixlnwxxitx{13.1}

\newcommand\pguavgpctnytfixlnwxxitx{1\%}

\newcommand\hyuavgnytfixlnwxxiax{13.2}

\newcommand\pguavgpctnytfixlnwxxiax{2\%}

\newcommand\hyuavgnytfixlnwxxibx{15.3}

\newcommand\pguavgpctnytfixlnwxxibx{5\%}

\newcommand\hyuavgnytctrynadjrx{5.0}
\newcommand\hywavgnytctrynadjrx{3.0}

\newcommand\hyuavgnytctrynadjtx{8.6}
\newcommand\hywavgnytctrynadjtx{5.6}

\newcommand\pguavgpctnytctrynadjtx{77\%}
\newcommand\pgwavgpctnytctrynadjtx{92\%}

\newcommand\hyuavgnytctrynadjax{8.6}
\newcommand\hywavgnytctrynadjax{5.6}

\newcommand\pguavgpctnytctrynadjax{75\%}
\newcommand\pgwavgpctnytctrynadjax{93\%}

\newcommand\hyuavgnytctrynadjbx{9.1}
\newcommand\hywavgnytctrynadjbx{5.8}

\newcommand\pguavgpctnytctrynadjbx{97\%}
\newcommand\pgwavgpctnytctrynadjbx{97\%}

\newcommand\hyuavgnytctrywxiitx{9.0}
\newcommand\hywavgnytctrywxiitx{5.8}

\newcommand\pguavgpctnytctrywxiitx{94\%}
\newcommand\pgwavgpctnytctrywxiitx{99\%}

\newcommand\hyuavgnytctrywxiiax{8.9}
\newcommand\hywavgnytctrywxiiax{5.7}

\newcommand\pguavgpctnytctrywxiiax{90\%}
\newcommand\pgwavgpctnytctrywxiiax{97\%}

\newcommand\hyuavgnytctrywxiibx{9.1}
\newcommand\hywavgnytctrywxiibx{5.8}

\newcommand\pguavgpctnytctrywxiibx{98\%}
\newcommand\pgwavgpctnytctrywxiibx{98\%}

\newcommand\hyuavgnytctrywxxitx{9.0}
\newcommand\hywavgnytctrywxxitx{5.7}

\newcommand\pguavgpctnytctrywxxitx{93\%}
\newcommand\pgwavgpctnytctrywxxitx{96\%}

\newcommand\hyuavgnytctrywxxiax{8.9}
\newcommand\hywavgnytctrywxxiax{5.7}

\newcommand\pguavgpctnytctrywxxiax{89\%}
\newcommand\pgwavgpctnytctrywxxiax{96\%}

\newcommand\hyuavgnytctrywxxibx{9.1}
\newcommand\hywavgnytctrywxxibx{5.8}

\newcommand\pguavgpctnytctrywxxibx{97\%}
\newcommand\pgwavgpctnytctrywxxibx{98\%}

\newcommand\hyuavgnytfnxlnnadjrx{3.4}

\newcommand\hyuavgnytfnxlnnadjtx{8.8}

\newcommand\pguavgpctnytfnxlnnadjtx{\textless1\%}

\newcommand\hyuavgnytfnxlnnadjax{9.2}

\newcommand\pguavgpctnytfnxlnnadjax{\textless1\%}

\newcommand\hyuavgnytfnxlnnadjbx{12.8}

\newcommand\pguavgpctnytfnxlnnadjbx{\textless1\%}

\newcommand\hyuavgnytfnxlnwxiitx{13.7}

\newcommand\pguavgpctnytfnxlnwxiitx{\textless1\%}

\newcommand\hyuavgnytfnxlnwxiiax{13.8}

\newcommand\pguavgpctnytfnxlnwxiiax{\textless1\%}

\newcommand\hyuavgnytfnxlnwxiibx{15.9}

\newcommand\pguavgpctnytfnxlnwxiibx{\textless1\%}

\newcommand\hyuavgnytfnxlnwxxitx{13.3}

\newcommand\pguavgpctnytfnxlnwxxitx{\textless1\%}

\newcommand\hyuavgnytfnxlnwxxiax{12.9}

\newcommand\pguavgpctnytfnxlnwxxiax{\textless1\%}

\newcommand\hyuavgnytfnxlnwxxibx{15.3}

\newcommand\pguavgpctnytfnxlnwxxibx{\textless1\%}

\newcommand\hyuavgnytfnwxiitx{11.6}

\newcommand\pguavgpctnytfnwxiitx{11\%}

\newcommand\hyuavgnytfnwxiiax{11.0}

\newcommand\pguavgpctnytfnwxiiax{10\%}

\newcommand\hyuavgnytfnwxiibx{14.1}

\newcommand\pguavgpctnytfnwxiibx{32\%}

\newcommand\hyuavgnytfnwxxitx{11.1}

\newcommand\pguavgpctnytfnwxxitx{7\%}

\newcommand\hyuavgnytfnwxxiax{10.4}

\newcommand\pguavgpctnytfnwxxiax{6\%}

\newcommand\hyuavgnytfnwxxibx{13.8}

\newcommand\pguavgpctnytfnwxxibx{28\%}

\newcommand\hyuavgnytnatlnadjrx{3.8}
\newcommand\hywavgnytnatlnadjrx{2.5}

\newcommand\hyuavgnytnatlnadjtx{8.2}
\newcommand\hywavgnytnatlnadjtx{5.2}

\newcommand\pguavgpctnytnatlnadjtx{66\%}
\newcommand\pgwavgpctnytnatlnadjtx{95\%}

\newcommand\hyuavgnytnatlnadjax{8.0}
\newcommand\hywavgnytnatlnadjax{5.1}

\newcommand\pguavgpctnytnatlnadjax{61\%}
\newcommand\pgwavgpctnytnatlnadjax{90\%}

\newcommand\hyuavgnytnatlnadjbx{8.9}
\newcommand\hywavgnytnatlnadjbx{5.4}

\newcommand\pguavgpctnytnatlnadjbx{97\%}
\newcommand\pgwavgpctnytnatlnadjbx{100\%}

\newcommand\hyuavgnytnatlwxiitx{8.8}
\newcommand\hywavgnytnatlwxiitx{5.4}

\newcommand\pguavgpctnytnatlwxiitx{90\%}
\newcommand\pgwavgpctnytnatlwxiitx{99\%}

\newcommand\hyuavgnytnatlwxiiax{8.8}
\newcommand\hywavgnytnatlwxiiax{5.3}

\newcommand\pguavgpctnytnatlwxiiax{90\%}
\newcommand\pgwavgpctnytnatlwxiiax{99\%}

\newcommand\hyuavgnytnatlwxiibx{8.9}
\newcommand\hywavgnytnatlwxiibx{5.4}

\newcommand\pguavgpctnytnatlwxiibx{98\%}
\newcommand\pgwavgpctnytnatlwxiibx{100\%}

\newcommand\hyuavgnytnatlwxxitx{8.8}
\newcommand\hywavgnytnatlwxxitx{5.4}

\newcommand\pguavgpctnytnatlwxxitx{91\%}
\newcommand\pgwavgpctnytnatlwxxitx{99\%}

\newcommand\hyuavgnytnatlwxxiax{8.7}
\newcommand\hywavgnytnatlwxxiax{5.3}

\newcommand\pguavgpctnytnatlwxxiax{88\%}
\newcommand\pgwavgpctnytnatlwxxiax{98\%}

\newcommand\hyuavgnytnatlwxxibx{8.9}
\newcommand\hywavgnytnatlwxxibx{5.4}

\newcommand\pguavgpctnytnatlwxxibx{98\%}
\newcommand\pgwavgpctnytnatlwxxibx{100\%}

\newcommand\hyuavgnytacrnwxiitx{11.0}
\newcommand\hywavgnytacrnwxiitx{8.7}

\newcommand\pguavgpctnytacrnwxiitx{\textless1\%}
\newcommand\pgwavgpctnytacrnwxiitx{75\%}

\newcommand\hyuavgnytacrnwxiiax{10.9}
\newcommand\hywavgnytacrnwxiiax{8.1}

\newcommand\pguavgpctnytacrnwxiiax{\textless1\%}
\newcommand\pgwavgpctnytacrnwxiiax{67\%}

\newcommand\hyuavgnytacrnwxiibx{13.9}
\newcommand\hywavgnytacrnwxiibx{9.5}

\newcommand\pguavgpctnytacrnwxiibx{\textless1\%}
\newcommand\pgwavgpctnytacrnwxiibx{91\%}

\newcommand\hyuavgnytacrnwxxitx{9.2}
\newcommand\hywavgnytacrnwxxitx{7.7}

\newcommand\pguavgpctnytacrnwxxitx{\textless1\%}
\newcommand\pgwavgpctnytacrnwxxitx{59\%}

\newcommand\hyuavgnytacrnwxxiax{9.7}
\newcommand\hywavgnytacrnwxxiax{7.8}

\newcommand\pguavgpctnytacrnwxxiax{\textless1\%}
\newcommand\pgwavgpctnytacrnwxxiax{61\%}

\newcommand\hyuavgnytacrnwxxibx{13.7}
\newcommand\hywavgnytacrnwxxibx{9.5}

\newcommand\pguavgpctnytacrnwxxibx{\textless1\%}
\newcommand\pgwavgpctnytacrnwxxibx{90\%}

\newcommand\hyuavgnytrgnnadjrx{4.1}
\newcommand\hywavgnytrgnnadjrx{3.1}

\newcommand\hyuavgnytrgnnadjtx{10.7}
\newcommand\hywavgnytrgnnadjtx{7.4}

\newcommand\pguavgpctnytrgnnadjtx{2\%}
\newcommand\pgwavgpctnytrgnnadjtx{53\%}

\newcommand\hyuavgnytrgnnadjax{10.2}
\newcommand\hywavgnytrgnnadjax{7.3}

\newcommand\pguavgpctnytrgnnadjax{1\%}
\newcommand\pgwavgpctnytrgnnadjax{51\%}

\newcommand\hyuavgnytrgnnadjbx{14.1}
\newcommand\hywavgnytrgnnadjbx{9.6}

\newcommand\pguavgpctnytrgnnadjbx{3\%}
\newcommand\pgwavgpctnytrgnnadjbx{88\%}

\newcommand\hyuavgnytrgnwxiitx{15.0}
\newcommand\hywavgnytrgnwxiitx{9.3}

\newcommand\pguavgpctnytrgnwxiitx{10\%}
\newcommand\pgwavgpctnytrgnwxiitx{81\%}

\newcommand\hyuavgnytrgnwxiiax{14.3}
\newcommand\hywavgnytrgnwxiiax{8.9}

\newcommand\pguavgpctnytrgnwxiiax{9\%}
\newcommand\pgwavgpctnytrgnwxiiax{75\%}

\newcommand\hyuavgnytrgnwxiibx{16.8}
\newcommand\hywavgnytrgnwxiibx{9.9}

\newcommand\pguavgpctnytrgnwxiibx{15\%}
\newcommand\pgwavgpctnytrgnwxiibx{94\%}

\newcommand\hyuavgnytrgnwxxitx{14.6}
\newcommand\hywavgnytrgnwxxitx{8.9}

\newcommand\pguavgpctnytrgnwxxitx{8\%}
\newcommand\pgwavgpctnytrgnwxxitx{75\%}

\newcommand\hyuavgnytrgnwxxiax{13.6}
\newcommand\hywavgnytrgnwxxiax{8.6}

\newcommand\pguavgpctnytrgnwxxiax{7\%}
\newcommand\pgwavgpctnytrgnwxxiax{71\%}

\newcommand\hyuavgnytrgnwxxibx{16.7}
\newcommand\hywavgnytrgnwxxibx{9.9}

\newcommand\pguavgpctnytrgnwxxibx{14\%}
\newcommand\pgwavgpctnytrgnwxxibx{93\%}

\title{Story Beyond the Eye: Glyph Positions Break PDF Text Redaction}
\author{Maxwell Bland}
\email{mb28@illinois.edu}
\author{Anushya Iyer}
\email{anushya2@illinois.edu}
\author{Kirill Levchenko}
\email{klevchen@illinois.edu}
\affiliation{%
  \institution{University of Illinois, Urbana-Champaign}
  \city{Urbana}
  \state{Illinois}
  \country{USA}}

\begin{document}
\begin{abstract}
In this work we find that many current redactions of PDF text are insecure due to non-redacted character positioning information.
In particular, subpixel-sized horizontal shifts in redacted and non-redacted characters can be recovered and used to effectively \emph{deredact} first and last names.
Unfortunately these findings affect redactions where the text underneath the black box is removed from the PDF.

We demonstrate these findings by performing a comprehensive vulnerability assessment of common PDF redaction types.
We examine 11 popular PDF redaction tools, including Adobe Acrobat, and find that they leak information about redacted text.
We also effectively deredact hundreds of real-world PDF redactions, including those found in OIG investigation reports and FOIA responses.

To correct the problem, we have released open source algorithms to fix trivial redactions and reduce the amount of information leaked by \emph{nonexcising} redactions (where the text underneath the redaction is copy-pastable).
We have also notified the developers of the studied redaction tools.
We have notified the Office of Inspector General, the Free Law Project, PACER, Adobe, Microsoft, and the US Department of Justice. 
We are working with several of these groups to prevent our discoveries from being used for malicious purposes.
\end{abstract}

\maketitle


\section{Introduction}
\label{sec:intro}
Redaction is a centuries-old process of removing or obscuring parts of a document to prevent their disclosure. 
In the past this was performed by black marker or by physically cutting parts of the document out with scissors.
However, with the move to digital, paperless representations of documents, the process of redaction is typically performed by software tools.
The process of removing information from digital documents is error-prone, as demonstrated by several high-profile examples~\cite{embarassingRedact}.
However, no existing work has considered the role the PDF file representation plays in redaction security.

In this work, we identify problems in PDF file text redaction.
It is well known that the width of the characters in a proportional-width font can leak information about redacted text.
However, we find PDF files include sub-pixel sized character position shifts, which can leak more information about redacted text than width alone.
Failing to understand these shifts exist leads to incorrect or imprecise deredaction. 
We are the first to identify this problem.

This led us to measure the severity of information leaks in PDF text redactions.
We discovered methods for breaking redactions occurring on documents processed by optical character recognition (OCR) software and find that \emph{rasterizing} a document (converting the PDF to an image) increases the security of redactions but does not remove information leaks.
We also find a typical PDF document authored in Microsoft Word leaks about 13 bits of information about a redacted surname. 
This is enough for attacker to consistently identify a single individual among 8,000 candidates (say, employees at a company or potential confidential informants in a gang), or $\approx$500 potential first initial, surname pairs from the 3.9 million possible given US census and Social Security Administration data.

We also consider \emph{nonexcising} redactions which do not remove any text from the document.
The text of these redactions can be copied and pasted from the source document.
Our study and techniques located thousands of nonexcising redactions in US court documents.
We notified both the US court system and the Free Law Project about these redactions.

We trace the source of the insecure redactions by examining 11 popular PDF redaction tools and find that two do not work at all.
These two tools leave the text underneath the selection marked for redaction in the PDF document.
The remaining tools, including those from Adobe Systems, remove the redacted text and replace it with a black rectangle.
Without additional defenses, this practice leaks significant information about the redacted text.

We built a tool, \maxray, which we use to identify, break, and fix redaction information leaks across millions of PDF files.
During the deredaction stage, \maxray\ allows an attacker to test which strings from a dictionary of candidates create the same PDF output as the original redacted PDF file.
We applied \maxray\ to study the impact of our findings on three publicly-available corpora of PDF documents.

We have notified organizations whose redacted documents we study and affected vendors of redaction tools and are working with several of these groups to remediate the problem.
To address the discovered vulnerabilities, we release tools for identifying and securing poorly redacted PDF text at [anonymized for peer review].

The contributions of this work are:
\begin{prettylist}
\item A survey of 11 popular PDF redaction tools. 
      We find two tools have redaction features which do not actually remove text marked for redaction, making the redacted document vulnerable to copy-paste attacks. 
      The remaining tools preserve glyph positioning information which can be used for glyph displacement attacks.
\item The discovery of redacted information leaks due to sub-pixel sized glyph position shifts.
      We provide a precise analysis of the role and impact of these shifts on the process of deredaction.
\item An information theoretic measurement of PDF redaction security.
      A redaction of a first name and surname from a voter registration database in a PDF produced by Microsoft Word's ``Save as PDF'' feature can leak up to 15.9 bits of information out of 19.6, meaning an adversary can correctly break such a redaction 38\% of the time.
\item Novel advisories regarding the security of PDF redactions and correct redaction methods.
      PDF redactions where the set of candidates may be reasonably constrained, e.g. a redaction of a politician's name, are not secure.
      Redactions of individuals' names where no other adjacent words are redacted are not secure.
\item A survey of millions of publicly-available redacted documents. 
      We find many of the documents provided by the US Courts, the Office of Inspector General (OIG), and Freedom of Information Act Requests (FOIA) are vulnerable.
      We found over 6,000 vulnerable nonexcising redactions. 
      We found over 300 vulnerable excising redactions.
  \item An extensive notification of affected parties and the release of several software tools for the development of advanced redaction defenses. 
        We have collaborated with the US government and third party software providers to correct and prevent the occurrence of discovered redaction vulnerabilities.
\end{prettylist}

This rest of this paper is organized as follows.
Section~\ref{sec:pdftext} gives background on how PDF text redactions leaks information.
Section~\ref{sec:tools} surveys PDF redaction tools' preservation of redacted information. 
Section~\ref{sec:maxray} describes \maxray's implementation. 
Section~\ref{sec:eval} measures mutual information shared by redacted and non-redacted PDF text.
Section~\ref{sec:wild} measures the number and diversity of redaction information leaks in public PDF documents.
Section~\ref{sec:disc} discusses defenses and future work.
Section~\ref{sec:related} addresses related work and Section~\ref{sec:concl} concludes.

\section{PDF Redaction Security}
\label{sec:pdftext}
\label{sec:textobj}

The security of PDF text redaction depends on the specification of the PDF document.
We consider two types of PDF documents.
One is a \raisebox{-3pt}{\includegraphics[width=0.75in]{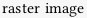}} of the original document.
The other PDF document type contains text data for both the font and the layout of each character (\emph{glyph}) on the page.

We focus our discussion on non-raster PDF documents, however, the concepts presented apply to raster PDF documents.
We consider the effect rasterization has on redaction information leaks when discussing it as a defense in Section~\ref{sec:defenses} and include a discussion of different document scanners in Appendix~\ref{sec:doc-scanners}.

\begin{figure}[h!]
\centering
\includegraphics[width=1.75in]{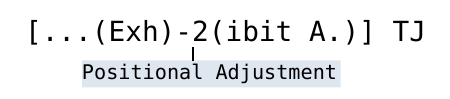}
    \caption{The TJ text showing operator, which specifies the glyphs to render and, by reference to a font object (not shown), their widths, along with any associated positional adjustments, given in text space units.}\label{fig:tj}
\end{figure}

PDF documents can render text in numerous ways, including by use of a text showing operator, one of which (TJ) is depicted in Figure~\ref{fig:tj}.
The TJ operator takes as arguments a string of text and a vector of \emph{positional adjustments} which displace the character with respect to a default position.
This default position is usually a fixed offset from the previous character equivalent to the \emph{advance width} of the previous character defined elsewhere in the PDF document.

For this paper, we converted the complex set of PDF text rendering operations into a uniform \emph{intermediate representation}, consisting of a minimal set of metrics (e.g. font size) and a series of TJ operators.\footnote{We have released code for performing this conversion at [anonymized for peer review].}
The intermediate representation presents PDF text as, effectively, a set of advance widths and \emph{glyph shifts} which are the sum of all the individual positioning operations applied to a glyph.
This conversion was necessary to account the large number of ways in which text can be rendered in PDF documents.\footnote{For example, in the case of a TJ operator, the actual \emph{glyph shift} includes any offset due to a positional adjustment as \emph{part} of the calculation of its value.}

The positional adjustment in Figure~\ref{fig:tj} is $-2$ \emph{text space units} between the \emph{h} and \emph{i} glyphs.
Text space units express glyph shifts, where 1,000 units almost always\footnote{PDF offers the ability to redefine the text space. \maxray\ accounts for this.} equals the point size of the font times \nicefrac{1}{72} of an inch.
For a 12-point font, 1 unit equals 1/6,000 of an inch (0.0042~mm).

Glyph advance widths and glyph shifts create a security concern.
Section~\ref{sec:tools} finds that most PDF redaction tools replace text selected for redaction with a single large shift of the same width as the redacted text showing operator, creating the two significant security risks:

\begin{itemize}
    \item The precise width of the redaction can be used to eliminate potential redacted texts (Sec.~\ref{sec:equiv-class}).
    \item Any non-redacted glyph shifts conditioned on redacted glyphs can be used to eliminate potential redacted texts (Sec.~\ref{sec:ms-word}).
\end{itemize}

Where appropriate in future sections, we also address concerns related to nonexcising redactions.
These redactions are cases where the text underneath the redaction can be selected and copied to the system clipboard from the PDF document.

\subsection{Glyph Shifts}
\label{sec:schemes}

The width of a PDF redaction depends on glyph shifts.
Without accounting for glyph shifts, redacted text guesses are imprecise and must account for error, reducing the potential of finding a unique match for redacted content.
See the discussion of rasterization error in Section~\ref{sec:defenses} for a precise measurement of how much shift error affects leaked redaction information.

The glyph shifts present in a PDF document are dependent on the specific \emph{workflow} used to produce the PDF document.
This includes an originating software, called the \emph{PDF producer} by the ISO 32000 PDF standard~\cite{pdfTwo}, and any software that may modify the PDF file contents thereafter, including, for example, a redaction tool.
A given workflow creates a specific pattern of glyph shifts, determining, in part, the security of any redacted text.
We identify two types of glyph shifting schemes:

\begin{itemize}
    \item \emph{Independent}: the glyph shifts for a given character are not dependent on any other character in the document in any way.
    \item \emph{Dependent}: the glyph shifts for a given character are dependent on some other character in the document in some way.
\end{itemize}

We call independent schemes \emph{unadjusted} when there are no shifts on any character.
Google Docs' \emph{Export to PDF} option produces an unadjusted scheme.

The frequency of documents with a given shift scheme varies by corpus (Sec.~\ref{sec:wild}).
For example, in our Freedom of Information Act (FOIA) corpus about 6\% of redacted text fragments are unadjusted, in contrast to 5.3\% across all corpora.

\subsubsection{Equivalence Classes.} 
\label{sec:equiv-class}
Before discussing these schemes further, we introduce the idea of width and shift equivalence classes.
A shift equivalence class is a set of lists of glyphs of the same length with identical shift values.
A width equivalence class is a set of glyphs and associated shifts with the same width.

\begin{table}\centering
\caption{Glyph width equivalence classes for the specific default version of
    Times New Roman used by Microsoft Word. Each number is the width in given to the
    glyph by the font file and each set of letters has equivalent widths.}
\label{tab:tnr-glyph-widths}
\small
\begin{tabular}{@{\quad}ll@{\quad\quad\quad}ll}
569 & ijlt & 1251 & ELTZ \\
683 & Ifr & 1366 & BCR \\
797 & Js & 1479 & ADGHKNOQUVXYw \\
909 & acez & 1593 & m \\
1024 & bdghknopquvxy & 1821 & M \\
1139 & FPS & 1933 & W
\end{tabular}
\end{table}

Table~\ref{tab:tnr-glyph-widths} gives an example of the width equivalence classes for glyphs in a Times New Roman font\footnote{
    Different versions of a font can exist. 
    We use the default versions available from Microsoft throughout this paper.
} without shifts, next to their widths as specified by the font file.
The glyphs \emph{I}, \emph{f}, and \emph{r} are exactly half the width of the glyphs of \emph{B}, \emph{C}, and \emph{R}. 
When typeset using Times New Roman, the words \emph{martian}, \emph{templar}, and \emph{mineral} all have the same width, as do the anagrams of those words \emph{tamarin}, \emph{trample}, and \emph{railmen}.

The PDF specification does not include any specific signifiers for redacted text.
However, residual specification information after redaction, such as glyph positions, can be used to reasonably rule out large numbers of candidate width and shift equivalence classes for redacted text.
None of the prior words in this paragraph are in the width equivalence class of the word \emph{cat}.

\subsubsection{Independent Schemes.}
\label{sec:gdocs}
\label{sec:adobe-ocr}

In an independent glyph shifting scheme, the security of a redaction may be considered dependent on the size of the width equivalence class indicated by the PDF document's residual glyph positioning information.
That is, the positions of glyphs prior to and succeeding the redaction may leak the width of redacted text.
The scheme's specific glyph shifts can make a given width equivalence class leak more or less redacted information by making width of individual glyphs more or less unique.

As an example, consider redacting a single letter \emph{l} as opposed to \emph{m} in the TNR scheme from Table~\ref{tab:tnr-glyph-widths}.
\emph{m} is the only glyph in its width equivalence class, so it may be possible to determine the letter \emph{m} was redacted uniquely.
Whereas if \emph{l} is redacted, then the residual information indicates the redacted letter could be any one of \emph{i}, \emph{j}, \emph{l}, or \emph{t}.
However, if \emph{l} were always accompanied by a glyph shift distinguishing it from these other three letters, then the scheme would leak more information.

We acknowledge that there is no \emph{guaranteed} correlation between glyph positions before and after redaction.
Redaction may reposition glyphs in such a way as to destroy accurate width information.
However, in Section~\ref{sec:tools} we find this is almost always the case for commonly accessible redaction tools.
We have also been informed that in some contexts there are (legal) restrictions on changing the glyphs or glyph positioning of a redacted document.

\emph{Scanned Documents.}
Many documents with independent shifting schemes are the result of an optical character recognition (OCR) process.\footnote{
We discuss physical document scanners in Appendix~\ref{sec:doc-scanners}.
}
This process embeds a non-raster representation of the document's text in the resulting PDF document.
This often allows the document's text to be searched and copied from by software tools.

We note that attacking the OCR overlay is not always as straightforward as attacking a PDF produced without OCR.
For example, attacking a redaction performed on the output of Adobe Acrobat Pro's OCR workflow requires using two side channels embedded in the PDF specification of the redacted document, detailed in Appendix~\ref{sec:acro-pro-deets}.
Using these side-channels, we find documents processed with OCR using Acrobat Pro and then subsequently redacted are vulnerable to the attacks presented in this paper.

\subsubsection{Dependent Schemes.}
\label{sec:ms-word}

A dependent scheme is more dangerous to the security of redacted text than an independent scheme.
In these schemes non-redacted glyph shifts can be dependent upon redacted glyph information, because the non-redacted glyph shifts can be determined \emph{before} redaction (Sec.~\ref{sec:tools}).

\emph{Microsoft Word ``Save as PDF''.}
In this work, we focus on a class of dependent schemes defined by the Microsoft Word software's \emph{Save As PDF} command.\footnote{
Shifting schemes generated by other workflows are detailed in Appendix Table~\ref{tab:pdfflows}.
}
We reverse-engineered the glyph shifting scheme produced by this command in Microsoft Word for Windows desktop versions 2007 to 2019.
This process took around several months. 
Word 2007 to 2016 use one scheme, and Word 2019 to present versions use another.
These two schemes affect thousands of real world document redactions (Sec.~\ref{sec:wild}).

\lstset{ %
language=C++,                
basicstyle=\ttfamily\footnotesize,       
numbers=left,                   
numberstyle=\footnotesize,      
stepnumber=1,                   
numbersep=5pt,                  
backgroundcolor=\color{white},  
showspaces=false,               
showstringspaces=false,         
showtabs=false,                 
frame=single,           
tabsize=2,          
captionpos=b,           
breaklines=true,        
breakatwhitespace=false,    
escapeinside={\%*}{*)}          
}

\begin{figure}
\begin{lstlisting}
for (int j = i + 1; j < vs->size(); j++) {
    t = ttfScaledWidths[j] / 1000;
    d = internalMSWordWidths[j] / internalMSWordFontSize;
    ttf += t;
    msWord += d;
    disp = ttf - msWord;
    if (
        ((disp > 0.003) || (disp < -0.003)) && 
        i != vs->size() - 1
    ) {
      int adj = disp * 1000 + 0.5;
      vs->setShift(j, adj);
      ttf = msWord = 0;
    } else {
      vs->setShift(j, 0);
    }
}
\end{lstlisting}
\caption{Snippet of reverse engineered code representing how Microsoft Word leaks redacted character information into non-redacted characters in a PDF document.
    }
\label{fig:msword-snippet}
\end{figure}

The studied dependent schemes accumulate a What You See Is What You Get (WYSIWYG) error measurement for each glyph from left to right across each line of text.
If redacted content is not removed from a Word document before running ``Save as PDF'', redacted glyph positioning information affects the accumulated error value.
\emph{Thus information about the content of a redaction is leaked into the shifts applied to non-redacted characters.}\footnote{Different text justifications can leak \emph{more} information. For an extended discussion see Appendix~\ref{sec:mic-word-deets}.}
A majority of the redaction tools from Section~\ref{sec:tools} preserved all non-redacted glyph shifts.

Figure~\ref{fig:msword-snippet} depicts this behavior.
Word's internal representation of the layout of characters for display purposes does not exactly match that of the TrueType Font (TTF) embedded in the PDF document.
Word corrects for this small error between the two formats through use of glyph shifts.
Suprisingly, these modifications are done independent of the user's screen resolution.

We note the internal widths used on line 3 of Figure~\ref{fig:msword-snippet} are determined by a loop with no overflow reset and the redacted information held by the accumulator \emph{is not zero} after a single shift is written.
We refer the interested reader to Appendix Figure~\ref{fig:wysiwyg-alg}: the ``pixelWidths'' variable in this figure is scaled by a constant to produce the ``internalMSWordWidths'' in Figure~\ref{fig:msword-snippet}.

Word's shifting scheme depends on the document's edit history. 
Changing a character inside of a Microsoft Word document splits the internal representation of the text fragment containing the character into two fragments. 
This fragmentation resets the accumulation of glyph width error and affects the shifts emitted for a line of text. 
\maxray\ accounts for this by considering all potential edits to redacted text.\footnote{
    We found edits have no significant impact on the efficacy of deredaction.
    The information leaked is \emph{at least} as much as an independent glyph shifting scheme.
    See Appendix~\ref{sec:mic-word-deets} for the editing algorithm.
    }

We validated our Word model on hundreds of lines of Wikipedia text rendered to PDF by Word, several manually-crafted test cases, and text fragments from redacted documents found in the wild.
We give detail the Microsoft Word shifting scheme algorithms further in Appendix~\ref{sec:mic-word-deets}. 

We have submitted a bug report reporting these findings to Microsoft and have received a response indicating they are aware of the problem.
While Microsoft is not \emph{liable} for this vulnerability, removing the leaks from new versions of Word will improve the security of future redacted content.

\section{Redaction Tools}
\label{sec:tools}

\begin{table}
  \centering
  \caption{Type of information leaked by redaction tools. Many tools leak redacted text width and glyph shift information. We include a star next to Adobe's tool. After discussion with Adobe, they claimed they do not use a raster approach. We found empirically this was not the case.}
  \small
  \label{tab:tools}
  \begin{tabular}{lll}
\toprule
\colname{Redaction tool (flow)} & 
\colname{Version} &
\colname{Leaked information}                                      \\
\midrule
PDFzorro                        & N/A          & Full Text \\
PDFzorro (lock)                 & N/A          & Raster Width \\
PDFescape Online                & N/A          & Full Text \\
PdfElement (all)                & 8.3.10       & Width and Shifts \\
Qoppa PDF Studio                & 2021.0.3     & Width and Shifts \\
FoxIt PDF Pro (all)             & 11.0.1       & Width and Shifts \\
Nitro PDF                       & 13.58.0      & Width and Shifts \\
Opentext Brava (all)            & 16.6.4.55    & Width                           \\
Rapid Redact                    & 2.3          & Raster Width \\
redactpdf.com                   & N/A          & Raster Width \\
Adobe (all)                     & 21.5         & Raster Width* \\
Adobe (postpone save)           & 21.5         & Width and Shifts \\
IText PDFSweep                  & 7.1.6        & Width \\
\bottomrule
\end{tabular}

\end{table}

We surveyed redaction tools to determine how much redacted text information they leak.
Our results are in Table~\ref{tab:tools}. 
To find these tools, we searched the web for the terms ``PDF redaction tool,'' ``redaction tool,'' and ``document redaction tool''. We then downloaded all redaction tools listed in the first five result pages. 
These tools represent what the typical user may encounter when searching for and using redaction software.

We find the redaction security vulnerabilities detailed by this paper are affected by redaction tools in one of four ways:

\paragraph{Full Text.}
The first family of tools draws a black box over text selected for redaction. The original PDF text object is still present and users can select and copy the text behind the black box. Two tools that fall into this category are listed as leaking \emph{full text} in Table~\ref{tab:tools}. We consider this to be a severe vulnerability and reported this to their developers (see Sec.~\ref{sec:disc}).

\paragraph{Width and Shifts.}
The second family of redaction tools, referred to as \emph{shift preserving}, attempts to perform redaction while maintaining the document's vector format. 
When applied to text objects, these tools replace the redacted text with a shift of equivalent width so that text outside the redaction remains at the same position as in the original document. 
We note the default behavior of all these tools is to draw a black box over the redacted area.\footnote{This is typically user configurable. However, there are also often legal and organizational requirements on the style of redactions.}
Table~\ref{tab:tools} lists these as leaking \emph{width and shifts}.

\paragraph{Width.}
The third family, Opentext Brava and IText PDFSweep, are listed as only leaking \emph{width}:

\emph{Opentext Brava} changes the underlying PDF font and shifts for non-redacted glyphs, however, the width of the original redaction is maintained to a close approximation and the number of characters in the redacted word leaks because every redacted character is replaced with a space character and an associated glyph shift. 

\emph{IText PDFSweep} considers the glyph shifts applied to each word, removes them, and applies their sum to the last glyph in the word before the redaction.
This has the effect of maintaining the exact width of redacted text but removing information relating to how a specific glyph in a given word was originally shifted.

\paragraph{Raster Width.}
The last family of tools, referred to as \emph{rasterizing tools}, starts by converting every document page into a raster image and then blacks out those pixels of the image that fall inside the area selected for redaction. Because pixels outside the redaction area are untouched, it is possible to recover the approximate width of redacted text from the spacing between glyphs on either side of the redaction box (Sec.~\ref{sec:defenses}). These tools are listed in Table~\ref{tab:tools} as leaking the \emph{raster width} of the redacted text.

\section{The \maxray\ Tool Suite}
\label{sec:maxray}

Following our analysis of redaction tools, it was necessary to develop the \maxray\ tool suite in order to better understand and fix the problem of deredaction.
\maxray\ automates the location, analysis, and protection of vulnerable redactions.


\textbf{Locating Redactions.}
\maxray\ adopts different approaches for identifying nonexcising and excising redactions.

For nonexcising redactions, \maxray\ first detects filled boxes drawn over text.
This algorithm has a large number of false positives as it does not consider whether the rectangles are actually covering text in a meaningful way.
A benefit of this approach is that it is fast: we use it as a first pass for identifying vulnerable nonexcising redactions in Section~\ref{sec:wild}.

To avoid false positives, \maxray\ then checks whether the pixels in the bounding box of each PDF glyph are all the same color.
This does not handle complex cases (e.g. using an image to redact text), but we did not find cases of these more complex styles of redaction in practice.

We manually validated our nonexcising redaction location algorithm on PDF documents from RECAP, a website hosting US court documents, and discovered a false positive rate of 4\%.
An additional 7\% of these documents properly changed the underlying text (e.g. to REDACTED) or contained unintended redactions (e.g. a black box covering non-sensitive text).
This resulted in \numRECAPredactionDocs\ documents with nonexcising redactions, many of which included entire paragraphs of text.\footnote{We provided a list of these PDFs and associated case names to the US Courts.}
The algorithm encounters false negatives when it cannot identify rectangular draw commands.
We did not encounter false negatives: redaction draw commands are rarely ambiguous enough to prevent detection.

For excising redactions, \maxray\ first identifies spaces larger than a single space character\footnote{The tool is tuned for English language redactions for a number of reasons, chiefly the fact that the English alphabet has proportional widths.} between each pair of words in a document.
It then analyzes the pixel color values between the words to identify a drawn rectangle.\footnote{
    We did not attempt to locate redactions without some visually signifying feature.
}

We evaluated our excising redaction location algorithm's accuracy on a random sample of 1,000 corpora pages from Section~\ref{sec:wild}.
We compared against Tim B. Lee's~\cite{timblee} prior work on redaction location and against manual identification.
Lee's method flags every black rectangle drawn as a redaction and has a high false positive rate for some classes of documents (around 30\% for FOIA and nearly 100\% for RECAP).
With respect to false positives and false negatives, our algorithm is equivalent to manual analysis when locating what we deem to be vulnerable redactions.

The above approaches present a novel contribution of the present work.
We collaborated with the Free Law Project on their tool x-ray~\cite{xrayTool}, for identifying redactions.
When we began collaboration, they communicated to us that there was no efficient tool with low false positives for identifying excising and nonexcising redactions.
We have created such a tool, however, we note the purpose of this paper was not to solve the problem of redaction location.

\textbf{Analyzing Redactions.}
\maxray\ analyzes the security of a given redaction by developing a information fingerprint (i.e. a hash of the width and shift equivalence classes) of all leaked glyph positioning information.
For each attempted guess at the redacted content, \maxray\ matches the expected information fingerprint for this guess to the fingerprint recovered from the source document.
Attempted guesses are selected from a \emph{dictionary}.

If the guess information fingerprint does not match, that guess is ruled out as a potential redacted content. 
\maxray's results are only as good as the dictionary. 
If the redacted text is not in the dictionary, \maxray\ may return incorrect or no results.
Deredaction does not \emph{discover} the redacted text.

To minimize the chances of not including the correct redacted content in our guess dictionary, in the following sections we use a dictionaries consisting of multiple variations of possible redacted terms.
We also performed extensive validation of our results wherever possible.

As a proof-of-concept tool, \maxray\ relies on a human user to select the proper dictionary. 
While this process could be automated using machine learning, we chose to do this step manually in order to exclude a potential source of error.

Running on an eight-core consumer laptop (a ThinkPad T420), \maxray\ eliminates around 80,000 guesses per second.


\textbf{Protecting Redactions.}
\maxray\ protects vulnerable PDF redactions by first locating the nonexcising redactions and removing their underlying text from the PDF.
We then adopt a user-configurable level of information excisement by allowing users to optionally remove all non-redacted glyph shifts\footnote{
    Recall even if the redaction width is changed these shifts can leak information.
} and optionally convert the font to a monospaced one, scaling the size to preserve readability.
To protect excising redactions, \maxray\ can round up the size of all spaces between two words to some width, $n \times w$, where $n$ is some number of characters and $w$ is width of a single character in the monospace font.
\maxray\ can also remove any rectangular draw commands from the PDF so that the width of the redaction cannot be recovered by examining the width of any graphical box drawn to represent the redaction.

While these changes may be unacceptable for many users, they guarantee the redactions' security in the general case.
Monospaced redactions leak less than five bits of information (Sec.~\ref{sec:eval}).
We discuss further defenses and recommendations in Section~\ref{sec:disc}.
We have made code for implementing the above protections open source.

\section{Measuring Leaked Information}
\label{sec:synth}
\label{sec:eval}

We next demonstrate exactly how dangerous glyph shift aware redaction attacks are.
We evaluate three types of glyph shifting schemes (Sec.~\ref{sec:pdftext}): a monospaced scheme, a scheme with independent (unadjusted) shifts,\footnote{
    We use an unadjusted scheme for simplicity: each font evaluated in this section can be considered its own independent scheme (Sec.~\ref{sec:pdftext}).
} and two schemes with dependent glyph shifts (MS Word schemes).

We can quantify the amount of information leaked by glyph positioning in information theoretic terms, as the \emph{mutual information} of the redacted text and the glyph positioning, measured in bits~\cite{cover-and-thomas}.
Mutual information measures the reduction in uncertainty about the redacted text from knowledge of glyph positions of the surrounding text (the \emph{fingerprint} or \emph{hash} of leaked information).

\subsection{Experiment Setup}
\label{sec:evalsetup}

\paragraph{Text corpus.}
We estimate the amount of leaked information by simulating redaction on text from the New York Times Annotated Corpus~\cite{nytCorp}.
The corpus contains articles written and published by the New York Times between January 1, 1987 and June 19, 2007.
The NYT corpus contains 1,855,658 documents in total, with 1,027,427,021 total tokens, 1,505,676 of which are unique.

We considered each word and string of punctuation to be a separate token, excluding contractions.
For example, a period and quotation mark (.'') ending a quotation would be a token. 
J. Doe is tokenized to ``J'', ``.'', and ``Doe''.
However, contractions such as ``we're'' are \emph{not} tokenized into \emph{we}, \emph{'}, and \emph{re}.

\paragraph{Dictionaries.} 
The amount of information leaked also depends on prior information about the redacted text. 
For example, if we know the redacted text is one of \neltln\ American surnames, then this redaction leaks at most $\log_2 \neltln \approx \hxln$ bits.
In our experiment, we model this prior information as a \emph{dictionary}, a set of strings from which we assume the redacted text is drawn.

Based on an examination of real redactions in Section~\ref{sec:wild}, we constructed 10 dictionaries.
We list the sources of these dictionaries in Appendix~\ref{sec:dictdetail}.

\begin{itemize}[nosep]
\item\textit{Str.} All strings of 3--16 characters in length starting with a uppercase or lowercase letter followed by lowercase letters.
\item\textit{Acrn.} All strings of 2--5 uppercase characters.
\item\textit{Word.} English words including some proper nouns.
\item\textit{Ctry.} Official and common names of countries.
\item\textit{Rgn.} Names of regions, a superset of \emph{Ctry}.
\item\textit{Natl.} Nationalities, demonyms, and adjectives of regions and nationalities, sourced from lists on Wikipedia.
\item\textit{FN.} American given (first) names.
\item\textit{LN.} American surnames (last names).
\item\textit{\fixlnname.} All combinations of a name initial followed by surname (\emph{LN}).
\item\textit{\fnxlnname.} All combinations of a given name (\emph{FN}) followed by a surname (\emph{LN}).
\item\textit{FNLN and FILN.} \emph{\fnxlnname} and \emph{\fixlnname} filtered to only include combinations of name and surname that appear in the voter registration databases of the three US states.
    North Carolina~\cite{ncVoterData}, Ohio~\cite{ohVoterData}, and Washington~\cite{waVoterData} were chosen based upon the availability of publicly accessible data.
\end{itemize}

\begin{table}
\centering
\caption{
    Dictionaries containing candidate texts used for evaluating deredaction. 
    Stop words are excluded from the statistics.
}\label{tab:dicts}
\small\begin{tabular}{lrr@{\hspace{3em}}*{1}{rr}}
\toprule
\emph{Dict.}
& \emph{Size}
& \emph{$H_{u}(X)$}
& \emph{NYT Occ.}
& \emph{$H_{e}(X)$}
\\
\midrule
\emph{Str} & \neltstr & \hxstr & \noccnytstr & \hxnytstr \\
\emph{Acrn} & \neltacrn & \hxacrn & \noccnytacrn & \hxnytacrn \\
\emph{Word} & \neltword & \hxword & \noccnytword & \hxnytword \\
\emph{Ctry} & \neltctry & \hxctry & \noccnytctry & \hxnytctry \\
\emph{Natl} & \neltnatl & \hxnatl & \noccnytnatl & \hxnytnatl \\
\emph{Rgn} & \neltrgn & \hxrgn & \noccnytrgn & \hxnytrgn \\
\emph{\fnname} & \neltfn & \hxfn & \noccnytfn & 10.3 \\
\emph{\lnname} & \neltln & \hxln & \noccnytln & 13.1 \\
  \emph{\filnname} & \num{1.6e6}      &   20.6  & 1,265,265 & 17.0 \\
\emph{\fixlnname} & \neltfiln & \hxfiln & \noccnytfiln & 17.0 \\
  \emph{\fnlnname} & \num{8.9e6}        &   23.1 & 3,650,063 & 19.6 \\  
\emph{\fnxlnname} & \neltfnln & \hxfnln & \noccnytfnln & 19.6 \\
\bottomrule
\end{tabular}

\end{table}

Table~\ref{tab:dicts} lists the dictionaries and their statistics.
\emph{Size} gives the number of entries in the dictionary. 
\emph{NYT Occ.} gives the number of occurrences of words from the given dictionary in the NYT corpus.
$H_u(X)$ is the uniformly distributed information theoretic entropy of the dictionary, given by $\log_2 \mathit{Size}$.
For dictionaries of individuals' names (those with \emph{FN} and \emph{FI} prefixes), $H_e(X)$ is the entropy of the empirical distribution of dictionary words in the voter registration databases.\footnote{
    We use the voter registration database frequency for individuals' names to avoid bias present when using the NYT corpus to estimate the frequency of a given name.
    For example, ``Al Gore'' is more frequent in the NYT corpus than the population.
}
For the other dictionaries $H_e(X)$ is the entropy of the empirical distribution of dictionary words in the NYT corpus.

The voter registration databases contained \num{1.7e7} names total.
Note that if this measure was used instead of \emph{NYT Occ.}, FN, LN, FILN, and FNLN would be identical to the population size.

Columns $H_u(X)$ and $H_e(X)$ are an \emph{upper bound} on the amount of information a document can leak about a redacted element of that dictionary.
If the amount of information leaked is equivalent to these numbers, then every word in the dictionary can be uniquely identified by leaked information when redacted.

Compared to the brute force approach of using \emph{all possible} name combinations, using voter registration databases makes deredaction highly precise.
However, we do not use \emph{FNLN} and \emph{FILN} in the evaluation of Section~\ref{sec:wild} because we do not want to bias our results by assuming what state the person (whose name was redacted) votes in or whether they are registered to vote.

\emph{We do not include strings of numbers in our dictionary list as the glyph advance width and effect on glyph shifts is almost always identical for digits, making them effectively monospace, even if typeset in a variable-width font.}

\paragraph{Workflow.} 
We simulate redaction in a PDF created with three shifting schemes from Section~\ref{sec:pdftext} (Unadjusted, Word 2007 and Word 2019 ``Save as PDF''), in 10 point Times New Roman, Calibri, and Arial, the three most common fonts in our document corpus (Sec.~\ref{sec:wild-setup}).
These fonts account for 71.6\% of lines in the 40,000 documents in Table~\ref{tab:wildres}.
We include Courier as an example of a monospaced font.
The simulated PDF, formatted for US Letter size paper, is left-justified with 1-inch margins, giving a 468 point line width.

We use a 10 point font size for our experiments as an upper bound on the amount of leaked information.
We report results for a 12 point font size in Appendix~\ref{sec:12ptfonts-appdx}. 
Because each 12 point line has fewer characters, the larger font size leaks a little less information overall.

\subsection{Experiment Procedure}
\label{sec:mutinfmethod}
We parameterize each redaction with experimental parameters of shifting scheme, font size, font, and dictionary.
We then perform the following steps:

\begin{enumerate}[nosep]
\item Choose some word from the corpus that occurs in the given dictionary.
\item Format a PDF document containing the chosen word and surrounding text from the corpus using the parameterized font and shifting scheme.
\item Replace the chosen word at the point with each word in the parameterized dictionary (either uniformly or according to the frequency distribution).
\item Redact the dictionary word by replacing it with a glyph shift equal to its width.
\item Record the leaked information fingerprint for this dictionary word.
\end{enumerate}

In summary, we simulated the redaction of all possible dictionary words at a sample of occurrences of dictionary words in the NYT corpus.
We then computed the amount of mutual information leaked by the redacted document $Y$ about the redacted word $X$ (See Appendix~\ref{appdx:mutinf}).

In addition to mutual information, we calculated the probability that the leaked information can be used to correctly guess the redacted word.
In the uniform distribution this is a random guess from the (typically quite small) set of matching candidate words, and in the frequency distribution we select the candidate with the highest frequency in either the NYT corpus (for \emph{Str}, \emph{Acrn}, \emph{Word}, \emph{Ctry}, \emph{Natl}, and \emph{Rgn}) or in the three state voter registration databases (for \emph{FN}, \emph{LN}, \emph{\fnxlnname}, \emph{\fixlnname}, \emph{FNLN}, and \emph{FILN}).

\begin{table*}
\centering
\caption{Number of bits leaked (left) and probability of a correct guess (right) for different shifting schemes in simulated redactions of the NYT corpus set in 10pt font.
    Refer to Table~\ref{tab:dicts} for the total number of bits of information present in the candidate dictionary.
    ``Probability correct guess'' refers to the likelihood of randomly selecting the redacted word given the (typically small) set of matching candidate texts.
    }
\label{tab:nyt10}
    \small\resizebox{\columnwidth}{!}{\newcommand\nytxunknown{\raisebox{1.25pt}{---}}
\begin{tabular}{l@{\hspace{0.785em}}l@{}r@{\hspace{3pt}}*{3}{@{\hspace{6pt}}*{3}{@{\hspace{1pt}}r@{\hspace{1pt}}}}@{\hspace{12pt}}*{3}{@{\hspace{6pt}}*{3}{@{\hspace{1pt}}r@{\hspace{2pt}}}}}
\toprule
& &
& \multicolumn{9}{c}{Leaked information (bits)}
& \multicolumn{9}{c}{Probability correct guess} \\
\cmidrule(lr){5-11}\cmidrule(lr){14-20}
\multicolumn{2}{l}{\emph{\textbf{Distr}}} & {Courier}\kern-8pt
& \multicolumn{3}{c}{{\tfontname}\hspace*{10pt}}
& \multicolumn{3}{c}{{\kern-6pt\afontname}}
& \multicolumn{3}{c}{{\kern-16pt\bfontname}}
& \multicolumn{3}{c}{{\tfontname}}
& \multicolumn{3}{c}{{\afontname}}
& \multicolumn{3}{c}{{\bfontname}}
\\
& \emph{Dict}
& Mo 
& {\nadjshortname} & {\wxiishortname} & {\wxxishortname}
& {\nadjshortname} & {\wxiishortname} & {\wxxishortname}
& {\nadjshortname} & {\wxiishortname} & {\wxxishortname}
& {\nadjshortname} & {\wxiishortname} & {\wxxishortname}
& {\nadjshortname} & {\wxiishortname} & {\wxxishortname}
& {\nadjshortname} & {\wxiishortname} & {\wxxishortname}
\\
\midrule
\multicolumn{6}{l}{\emph{\textbf{\uniname}}}
\\
& \emph{\strname}
& \hyunytstrnadjrx
& \hyunytstrnadjtx & \nytxunknown & \nytxunknown
& \hyunytstrnadjax & \nytxunknown & \nytxunknown
& \hyunytstrnadjbx & \nytxunknown & \nytxunknown
& \pgupctnytstrnadjtx & \nytxunknown & \nytxunknown
& \pgupctnytstrnadjax & \nytxunknown & \nytxunknown
& \pgupctnytstrnadjbx & \nytxunknown & \nytxunknown
\\
& \emph{\acrnname}
& \hyunytacrnnadjrx
& \hyunytacrnnadjtx & \hyuavgnytacrnwxiitx & \hyuavgnytacrnwxxitx
& \hyunytacrnnadjax & \hyuavgnytacrnwxiiax & \hyuavgnytacrnwxxiax
& \hyunytacrnnadjbx & \hyuavgnytacrnwxiibx & \hyuavgnytacrnwxxibx
& \pgupctnytacrnnadjtx & \pguavgpctnytacrnwxiitx & \pguavgpctnytacrnwxxitx
& \pgupctnytacrnnadjax & \pguavgpctnytacrnwxiiax & \pguavgpctnytacrnwxxiax
& \pgupctnytacrnnadjbx & \pguavgpctnytacrnwxiibx & \pguavgpctnytacrnwxxibx
\\
& \emph{\wordname}
& \hyunytwordnadjrx
& \hyunytwordnadjtx & \hyuavgnytwordwxiitx & \hyuavgnytwordwxxitx
& \hyunytwordnadjax & \hyuavgnytwordwxiiax & \hyuavgnytwordwxxiax
& \hyunytwordnadjbx & \hyuavgnytwordwxiibx & \hyuavgnytwordwxxibx
& \pgupctnytwordnadjtx & \pguavgpctnytwordwxiitx & \pguavgpctnytwordwxxitx
& \pgupctnytwordnadjax & \pguavgpctnytwordwxiiax & \pguavgpctnytwordwxxiax
& \pgupctnytwordnadjbx & \pguavgpctnytwordwxiibx & \pguavgpctnytwordwxxibx
\\
& \emph{\ctryname}
& \hyuavgnytctrynadjrx
& \hyuavgnytctrynadjtx & \hyuavgnytctrywxiitx & \hyuavgnytctrywxxitx
& \hyuavgnytctrynadjax & \hyuavgnytctrywxiiax & \hyuavgnytctrywxxiax
& \hyuavgnytctrynadjbx & \hyuavgnytctrywxiibx & \hyuavgnytctrywxxibx
& \pguavgpctnytctrynadjtx & \pguavgpctnytctrywxiitx & \pguavgpctnytctrywxxitx
& \pguavgpctnytctrynadjax & \pguavgpctnytctrywxiiax & \pguavgpctnytctrywxxiax
& \pguavgpctnytctrynadjbx & \pguavgpctnytctrywxiibx & \pguavgpctnytctrywxxibx
\\
& \emph{\rgnname}
& \hyuavgnytrgnnadjrx
& \hyuavgnytrgnnadjtx & \hyuavgnytrgnwxiitx & \hyuavgnytrgnwxxitx
& \hyuavgnytrgnnadjax & \hyuavgnytrgnwxiiax & \hyuavgnytrgnwxxiax
& \hyuavgnytrgnnadjbx & \hyuavgnytrgnwxiibx & \hyuavgnytrgnwxxibx
& \pguavgpctnytrgnnadjtx & \pguavgpctnytrgnwxiitx & \pguavgpctnytrgnwxxitx
& \pguavgpctnytrgnnadjax & \pguavgpctnytrgnwxiiax & \pguavgpctnytrgnwxxiax
& \pguavgpctnytrgnnadjbx & \pguavgpctnytrgnwxiibx & \pguavgpctnytrgnwxxibx
\\
& \emph{\natlname}
& \hyuavgnytnatlnadjrx
& \hyuavgnytnatlnadjtx & \hyuavgnytnatlwxiitx & \hyuavgnytnatlwxxitx
& \hyuavgnytnatlnadjax & \hyuavgnytnatlwxiiax & \hyuavgnytnatlwxxiax
& \hyuavgnytnatlnadjbx & \hyuavgnytnatlwxiibx & \hyuavgnytnatlwxxibx
& \pguavgpctnytnatlnadjtx & \pguavgpctnytnatlwxiitx & \pguavgpctnytnatlwxxitx
& \pguavgpctnytnatlnadjax & \pguavgpctnytnatlwxiiax & \pguavgpctnytnatlwxxiax
& \pguavgpctnytnatlnadjbx & \pguavgpctnytnatlwxiibx & \pguavgpctnytnatlwxxibx
\\
& \emph{\fnname}
& \hyunytfnnadjrx
& \hyunytfnnadjtx & \hyuavgnytfnwxiitx & \hyuavgnytfnwxxitx
& \hyunytfnnadjax & \hyuavgnytfnwxiiax & \hyuavgnytfnwxxiax
& \hyunytfnnadjbx & \hyuavgnytfnwxiibx & \hyuavgnytfnwxxibx
& \pgupctnytfnnadjtx & \pguavgpctnytfnwxiitx & \pguavgpctnytfnwxxitx
& \pgupctnytfnnadjax & \pguavgpctnytfnwxiiax & \pguavgpctnytfnwxxiax
& \pgupctnytfnnadjbx & \pguavgpctnytfnwxiibx & \pguavgpctnytfnwxxibx
\\
& \emph{\lnname}
& \hyunytlnnadjrx
& \hyunytlnnadjtx & \hyuavgnytlnwxiitx & \hyuavgnytlnwxxitx
& \hyunytlnnadjax & \hyuavgnytlnwxiiax & \hyuavgnytlnwxxiax
& \hyunytlnnadjbx & \hyuavgnytlnwxiibx & \hyuavgnytlnwxxibx
& \pgupctnytlnnadjtx & \pguavgpctnytlnwxiitx & \pguavgpctnytlnwxxitx
& \pgupctnytlnnadjax & \pguavgpctnytlnwxiiax & \pguavgpctnytlnwxxiax
& \pgupctnytlnnadjbx & \pguavgpctnytlnwxiibx & \pguavgpctnytlnwxxibx
\\
& \emph{\filnname}
 & 2.9 & 8.6 & 13.3 & 12.6 & 8.7 & 13.0 & 12.2 & 13.0 & 15.6 & 15.5 & <1\% & 4\% & 2\% & <1\% & 4\% & 2\% & 2\% & 11\% & 11\%
\\
& \emph{\fixlnname}
& \hyuavgnytfixlnnadjrx
& \hyuavgnytfixlnnadjtx & \hyuavgnytfixlnwxiitx & \hyuavgnytfixlnwxxitx
& \hyuavgnytfixlnnadjax & \hyuavgnytfixlnwxiiax & \hyuavgnytfixlnwxxiax
& \hyuavgnytfixlnnadjbx & \hyuavgnytfixlnwxiibx & \hyuavgnytfixlnwxxibx
& \pguavgpctnytfixlnnadjtx & \pguavgpctnytfixlnwxiitx & \pguavgpctnytfixlnwxxitx
& \pguavgpctnytfixlnnadjax & \pguavgpctnytfixlnwxiiax & \pguavgpctnytfixlnwxxiax
& \pguavgpctnytfixlnnadjbx & \pguavgpctnytfixlnwxiibx & \pguavgpctnytfixlnwxxibx
\\
& \emph{\fnlnname}
 & 3.2 & 9.4 & 14.5 & 14.1 & 10.0 & 14.9 & 14.0 & 13.5 & 16.7 & 16.5 & <1\% & 3\% & 2\% & <1\% & 4\% & 3\% & 3\% & 8\% & 7\%
\\
& \emph{\fnxlnname}
& \hyuavgnytfnxlnnadjrx
& \hyuavgnytfnxlnnadjtx & \hyuavgnytfnxlnwxiitx & \hyuavgnytfnxlnwxxitx
& \hyuavgnytfnxlnnadjax & \hyuavgnytfnxlnwxiiax & \hyuavgnytfnxlnwxxiax
& \hyuavgnytfnxlnnadjbx & \hyuavgnytfnxlnwxiibx & \hyuavgnytfnxlnwxxibx
& \pguavgpctnytfnxlnnadjtx & \pguavgpctnytfnxlnwxiitx & \pguavgpctnytfnxlnwxxitx
& \pguavgpctnytfnxlnnadjax & \pguavgpctnytfnxlnwxiiax & \pguavgpctnytfnxlnwxxiax
& \pguavgpctnytfnxlnnadjbx & \pguavgpctnytfnxlnwxiibx & \pguavgpctnytfnxlnwxxibx
\\
\multicolumn{6}{l}{\emph{\textbf{Text frequency distr.}}}
\\
& \emph{\strname}
& \hywnytstrnadjrx
& \hywnytstrnadjtx & \nytxunknown  & \nytxunknown
& \hywnytstrnadjax & \nytxunknown  & \nytxunknown
& \hywnytstrnadjbx & \nytxunknown  & \nytxunknown
& \pgwpctnytstrnadjtx & \nytxunknown & \nytxunknown
& \pgwpctnytstrnadjax & \nytxunknown & \nytxunknown
& \pgwpctnytstrnadjbx & \nytxunknown & \nytxunknown
\\
& \emph{\acrnname}
& \hywnytacrnnadjrx
& \hywnytacrnnadjtx & \hywavgnytacrnwxiitx & \hywavgnytacrnwxxitx
& \hywnytacrnnadjax & \hywavgnytacrnwxiiax & \hywavgnytacrnwxxiax
& \hywnytacrnnadjbx & \hywavgnytacrnwxiibx & \hywavgnytacrnwxxibx
& \pgwpctnytacrnnadjtx & \pgwavgpctnytacrnwxiitx & \pgwavgpctnytacrnwxxitx
& \pgwpctnytacrnnadjax & \pgwavgpctnytacrnwxiiax & \pgwavgpctnytacrnwxxiax
& \pgwpctnytacrnnadjbx & \pgwavgpctnytacrnwxiibx & \pgwavgpctnytacrnwxxibx
\\
& \emph{\wordname}
& \hywnytwordnadjrx
& \hywnytwordnadjtx & \hywavgnytwordwxiitx & \hywavgnytwordwxxitx
& \hywnytwordnadjax & \hywavgnytwordwxiiax & \hywavgnytwordwxxiax
& \hywnytwordnadjbx & \hywavgnytwordwxiibx & \hywavgnytwordwxxibx
& \pgwpctnytwordnadjtx & \pgwavgpctnytwordwxiitx & \pgwavgpctnytwordwxxitx
& \pgwpctnytwordnadjax & \pgwavgpctnytwordwxiiax & \pgwavgpctnytwordwxxiax
& \pgwpctnytwordnadjbx & \pgwavgpctnytwordwxiibx & \pgwavgpctnytwordwxxibx
\\
& \emph{\ctryname}
& \hywavgnytctrynadjrx
& \hywavgnytctrynadjtx & \hywavgnytctrywxiitx & \hywavgnytctrywxxitx
& \hywavgnytctrynadjax & \hywavgnytctrywxiiax & \hywavgnytctrywxxiax
& \hywavgnytctrynadjbx & \hywavgnytctrywxiibx & \hywavgnytctrywxxibx
& \pgwavgpctnytctrynadjtx & \pgwavgpctnytctrywxiitx & \pgwavgpctnytctrywxxitx
& \pgwavgpctnytctrynadjax & \pgwavgpctnytctrywxiiax & \pgwavgpctnytctrywxxiax
& \pgwavgpctnytctrynadjbx & \pgwavgpctnytctrywxiibx & \pgwavgpctnytctrywxxibx
\\
& \emph{\rgnname}
& \hywavgnytrgnnadjrx
& \hywavgnytrgnnadjtx & \hywavgnytrgnwxiitx & \hywavgnytrgnwxxitx
& \hywavgnytrgnnadjax & \hywavgnytrgnwxiiax & \hywavgnytrgnwxxiax
& \hywavgnytrgnnadjbx & \hywavgnytrgnwxiibx & \hywavgnytrgnwxxibx
& \pgwavgpctnytrgnnadjtx & \pgwavgpctnytrgnwxiitx & \pgwavgpctnytrgnwxxitx
& \pgwavgpctnytrgnnadjax & \pgwavgpctnytrgnwxiiax & \pgwavgpctnytrgnwxxiax
& \pgwavgpctnytrgnnadjbx & \pgwavgpctnytrgnwxiibx & \pgwavgpctnytrgnwxxibx
\\
& \emph{\natlname}
& \hywavgnytnatlnadjrx
& \hywavgnytnatlnadjtx & \hywavgnytnatlwxiitx & \hywavgnytnatlwxxitx
& \hywavgnytnatlnadjax & \hywavgnytnatlwxiiax & \hywavgnytnatlwxxiax
& \hywavgnytnatlnadjbx & \hywavgnytnatlwxiibx & \hywavgnytnatlwxxibx
& \pgwavgpctnytnatlnadjtx & \pgwavgpctnytnatlwxiitx & \pgwavgpctnytnatlwxxitx
& \pgwavgpctnytnatlnadjax & \pgwavgpctnytnatlwxiiax & \pgwavgpctnytnatlwxxiax
& \pgwavgpctnytnatlnadjbx & \pgwavgpctnytnatlwxiibx & \pgwavgpctnytnatlwxxibx
\\
& \emph{\fnname}
 & 2.5 & 7.1 & 9.3 & 9.0 & 7.2 & 8.8 & 8.6 & 9.7 & 10.0 & 9.9 & 45\% & 79\% & 74\% & 46\% & 71\% & 68\% & 87\% & 93\% & 92\%
\\
& \emph{\lnname}
 & 2.7 & 7.8 & 10.8 & 10.4 & 7.9 & 10.7 & 10.1 & 11.4 & 12.2 & 12.1 & 28\% & 59\% & 53\% & 28\% & 58\% & 51\% & 66\% & 81\% & 79\%
\\
& \emph{\filnname}
 & 2.7 & 8.4 & 12.4 & 11.9 & 8.5 & 12.2 & 11.5 & 12.6 & 14.4 & 14.3 & 8\% & 30\% & 22\% & 8\% & 25\% & 21\% & 34\% & 53\% & 52\%
\\
& \emph{\fnlnname}
 & 3.1 & 9.2 & 13.8 & 13.6 & 9.8 & 14.3 & 13.5 & 13.2 & 15.9 & 15.8 & 4\% & 20\% & 19\% & 6\% & 24\% & 19\% & 16\% & 38\% & 37\%
\\
\bottomrule
\end{tabular}
}
\end{table*}

\subsection{Results}
Table~\ref{tab:nyt10} reports on the vulnerability of excising redactions.
The top half of the table shows the case where dictionary elements are drawn uniformly at random in step 3 above and the bottom half shows the case where dictionary elements are drawn according to their frequency of occurrence in the NYT corpus or, in the case of names, in the three states' voter registration databases. 
We have omitted rows for the \emph{\fnxlnname} and \emph{\fixlnname} dictionaries from the bottom half because they are identical to the \emph{FNLN} and \emph{FILN} results, respectively.\footnote{
We considered using the empirical distribution generated by the independent distributions of first and last names in the voter registration databases.
However, the distribution of first names is \emph{not independent} of the distribution of surnames due to cultural naming conventions. 
}

The left side of the table shows the mutual information of the dictionary and the resulting document.
This is the number of bits of information leaked by the document about the redacted word.
These should be compared to the total entropy of the corresponding dictionary given in Table~\ref{tab:dicts}.

The right side of the table gives the probability correctly guessing the redacted text using only the information available after redaction.
This corresponds to success probability of a game in which a player knows the dictionary and tries to guess the redacted word based information in the redacted document.
In the non-uniform case, the optimal strategy is to guess the most likely (the highest occurrence frequency) word or phrase that produces the same redacted document.

The table gives statistics for four fonts: Courier, Times New Roman, Arial, and Calibri.
For each font, we show the amount of information leaked by an unadjusted shifting scheme (\nadjshortname), text produced using Word 2007--2016 (\wxiishortname), and Word 2019--2021 (\wxxishortname) dependent glyph shifting schemes. 
For Courier, we omit the \wxiishortname\ and \wxxishortname\ columns as monospaced fonts behave identically in the unadjusted and Word cases.

Table~\ref{tab:nyt10} does not show results for the \emph{Str} dictionary under the uniform distribution using Word schemes because simulating redaction of all \neltstr\ strings is prohibitively expensive. 
Results for the unadjusted positioning scheme were obtained without simulating redaction by exploiting the regular structure of the dictionary.

We compare redaction vulnerabilities across three categories, using the example of redacting a surname set in 10 point Times New Roman font to guide the reader:

    \textbf{Monospace (Mo).} For monospaced font redactions, the residual information in the document after redaction reveals the number of characters in the redacted word. 
The Courier column in Table~\ref{tab:nyt10} thus tells us how much information is revealed by knowing the number of letters in the word.
(For monospace fonts, which are always unadjusted, this is just the entropy of the distribution of word lengths.)
For example, knowing only the number of characters in a surname leaks 2.9 bits of information (out of 17.2) when guessing uniformly at random from the candidate redacted texts.
This has a $<1\%$ probability of success---redactions of monospace fonts are relatively secure.

\textbf{Unadjusted (Un).} For independent shifting scheme redactions, e.g. a PDF produced by Google Docs, the residual width of the redaction leaks more information than the number of characters redacted.
The width of a redaction of a surname (the \emph{\nadjshortname} row) with no glyph shifts provides 8.2 bits of information about the surname if it is chosen uniformly at random and 7.8 bits (out of 13.1) if it is chosen according to an empirical distribution. 
While the uniform distribution still has a $<1\%$ probability of success, the empirical distribution has a $28\%$ probability of success.

\textbf{Dependent (W07/W19).} For dependent shifting scheme redactions, the residual information after redaction leaks both width and shift equivalence classes (Sec.~\ref{sec:equiv-class}) and therefore more information.
If we redact a surname from a PDF produced by Word 2007--2016, the resulting document leaks 12.4 bits of information about a name chosen uniformly at random and 10.8 bits when chosen according to the empirical distribution given by voter registration databases.
With the inclusion of these information leaks, the probability of a correct guess under the uniform distribution is $11\%$ and the probability under an empirical distribution is greater than $50\%$.

We found leaks of up to 15 bits of information about redacted text in dependent (Microsoft Word ``Save as PDF'') glyph shifting schemes.
These schemes present a significant security concern for excising redactions.
Even without considering the frequency of names in the population, single word redactions have a greater than 5\% chance of being broken.
When an adversary can use statistical likelihoods of names, two word redactions of first names and surnames face a 1 in 5 chance of being broken, and an adversary's chances are only better for other fonts.

Calibri, the default font in Microsoft Office since 2007, leaks the most information because the font has a greater variety of character widths than Times New Roman or Arial. 
Recall that the empirical entropy of the surname dictionary is 13.1 bits (Table~\ref{tab:dicts}), so a redacted surname set in 10-point Calibri using a Word 2007 shifting scheme leaks almost all the information available and can be correctly deredacted in roughly 81\% of cases.

\section{Redaction in the Wild}
\label{sec:wild}

We next consider the question of whether these findings are applicable to real documents.
This is not immediately clear because glyph shifts may be modified by a variety of software workflows.\footnote{
    For example, by opening the PDF produced by word \emph{and then} modifying it using Adobe Acrobat. 
    We perform an analysis of a few different workflows in Appendix~\ref{appdx:flows}.
}
It is also unclear whether there exist a significant number of vulnerable redactions in real documents.

We consider both \emph{nonexcising} and \emph{excising} redactions in this section.
As mentioned in Section~\ref{sec:intro}, nonexcising redactions retain redacted text in the PDF and only \emph{visually} obscure the text.
Excising redactions, on the other hand, remove the text from the PDF file.

We chose to study redactions of names (e.g. first names, surnames, and country names) in this section because they are the most common, discussed further below, and because their release presents a privacy concern.
We do not release any of these names and have taken steps to notify affected parties in order to ensure no individual will be adversely affected by the present analysis (Section~\ref{sec:ethics}).

\subsection{Experiment Setup}
\label{sec:wild-setup}

\paragraph{Corpora.} We use the following real world document corpora: 

\begin{enumerate}
    \item \emph{FOIA.} Documents obtained via the US Freedom of Information Act (FOIA) on governmentattic.org~\cite{govattic}. 
This corpus provides us with independently selected documents with some public interest.
    \item \emph{OIG.} Office of the Inspector General (OIG) reports hosted by oversight.gov~\cite{oigReports}. 
The OIG is a US Government oversight branch tasked with preventing unlawful operation of other government branches.
This corpus allowed us to measure the impact deredaction may have on documents from a high-profile and large organization.
    \item \emph{DNSA.} Digital National Security Archive (DNSA) documents produced after 2010~\cite{dnsaSite}. 
The DNSA is a set of historical US government documents curated by scholars. 
That is, we found redaction information leaks affect significant historical documents.
    \item \emph{RECAP.} CourtListener's RECAP court document archive.
RECAP mirrors PACER, the US Federal Courts' docketing system~\cite{pacerSite}, and contains over 10 million documents.
We use RECAP to measure the impact of nonexcising redactions (discussed below).
    \item \emph{rRECAP} the subset of RECAP documents returned for the search string ``redacted''.
We chose to include rRECAP because running the excising redaction location algorithm mentioned in Section~\ref{sec:maxray} on the entire RECAP corpus would be both computationally and financially prohibitive.
\end{enumerate}

Only the RECAP corpus contained nonexcising redactions and our results for this corpus are reported with respect to nonexcising redactions.
Our results for all other corpora are reported with respect to excising redactions.

\paragraph{Dictionaries.}
We chose to restrict our evaluation to first name and last name redactions, which are very common in our corpus.
In a sample of 100 redactions from our corpora, 52 were personal names (determined manually based on the surrounding text), 26 were multi-word phrases too long to attack, 17 were numbers (not vulnerable to attack), 3 were pronouns (trivial to attack), and 2 did not have identifiable semantics.
We composed a dictionary by taking the union of the first name and last name sets introduced in Section~\ref{sec:eval}.

We also include titles, such as ``Mr.'' and ``Mrs.'', as well as initials, such as ``J.'' and ``S.'', in our dictionary.
However, our results for matches \emph{remove} these prefixes, so ``Ms. Doe'', ``J. Doe'', and ``Doe'' count as a single match.

Our final dictionary contained \nameDictSizeComb\ entries.
We do not include \emph{\fnxlnname} as testing this dictionary takes 6 hours on an Intel Xeon Silver 4208, 2.10 GHz, 32-core server and the result sets are typically large.

We exclude the DNSA from this dictionary restriction. 
The redactions matching the Microsoft Word dependent glyph shifting scheme in the DNSA were of an acronym, demonyms, and countries.
We used these dictionaries when performing deredaction on the DNSA (see Appdx.~\ref{sec:dictdetail}).

\paragraph{Workflow.}

In this section we evaluate a redaction if either:

\begin{enumerate}[nosep]
    \item The redacted text is present in the PDF (vulnerable to copy-paste attack); or
    \item The redacted text is not present, but the document retains glyph shifting scheme information where:
        \begin{itemize}[nosep]
            \item The scheme matches a Word ``Save as PDF'' shifting scheme,
            \item The redaction appears to be a name, e.g. ``Jane'', and
            \item The redaction is the first from left to right on the line of text.
        \end{itemize}
\end{enumerate}

These criteria provide a uniform and accurate proof-of-concept evaluation of the impact of the discovered redaction vulnerabilities.

\paragraph{Limitations.}
We chose to restrict our attacks to the Microsoft Word schemes because these leak the most information, and are thus of the greatest concern.
Other shifting schemes exist and evaluating these schemes' security will require further reverse engineering.
Cases matching the Microsoft Word ``Save as PDF'' workflow represented 8.8\% of redaction instances across all four corpora.


Section~\ref{sec:pdftext} noted Microsoft Word's glyph shifting scheme leaks redacted information from left to right. 
We therefore considered only the first redaction on a line to remove any dependence on prior correct guesses, which would be necessary to attack the second redaction on a line.

The cases of redaction we report are interesting and chosen conservatively.
In reality, there may be additional vulnerable redactions not included in our count.

\subsection{Experiment Procedure}
\label{sec:adjschmident}
\label{sec:ident-weak}

We take three steps to ensure the results of our experiments are accurate:
\begin{enumerate}
    \item We require all potential deredactions to exactly match all present PDF glyph positioning information.
    \item For excising redactions, it is necessary to identify the shifting scheme used by the PDF document.
This process avoids incorrect matching and filters out regions of documents with idiosyncratic formatting.
For example, Microsoft Word documents may have alternative layouts for text, e.g. text boxes, tables, and line numbers. 

In the present analysis, we consider a PDF page to \emph{match} a scheme if we identify greater than 100 glyphs with shift values matching the scheme on the page.
We only count entire lines: all of a line's glyph positioning information must match the scheme exactly to count toward the 100 glyph threshold.
\item We manually classified each excising redaction as being a redacted name based upon the content of the surrounding text.
We were conservative with our classification and only attack redactions we are \emph{certain} are of names.
\item Where possible, we validated our findings using public information.
\end{enumerate}

Although not critical to our results, we also identified redactions in documents using unadjusted and Adobe OCR glyph shifting schemes.
In Section~\ref{sec:eval}, we found these schemes leak 2--3 bits less information than documents produced by Microsoft Word.

To make sure we were missing any PDF documents originating from Microsoft Word, we also counted redactions within a 10 text space unit $L_{1}$ distance from our Microsoft Word model.
These PDFs were the result of a non-standard workflow, e.g. creation using Word 365, the web interface for Microsoft Word.
In our results, we label these redactions as \emph{Near Word}.

\begin{table}
  \centering
    \caption{Top: Glyph shifting schemes identified in redacted corpora pages. Bottom: Deredaction results for names tagged.}
\label{tab:wildres}
\small
\begin{tabular}{lrr@{\hspace{3pt}}r@{\hspace{3pt}}r|r}
  \toprule
    Metric             & FOIA              & OIG              & DNSA                & rRECAP               & RECAP                     \\
  \midrule
Documents          & 3,145             & \oigNumDocs           & \dnsaNumDocs        & \recapNumDocs        & $\approx10^{7}$           \\
  Pages              & \foiaPg           & \oigPg          & \dnsaPg             & \recapPg             & \numRECAPpages            \\
  Redacted PDFs      & 236               & 1255             & \dnsaNumDocsWRedact & \recapNumDocsWRedact & \numRECAPredactionDocs    \\
  Redactions         & \foiaMetaRedact & \oigMetaRedact & \dnsaMetaRedact     & \recapMetaRedact     & \numRECAPNameredactions   \\
  \ Unadjusted     & 2,844             & 314              & \unadjDNSA          & \unadjRECAP          & \numRECAPnadj             \\
  \ Adobe OCR      & 3,406             & 1,814            & \ocrDNSA            & \ocrRECAP            & \numRECAPocr              \\
  \ Near Word   & 175               & 114              & \mswDNSANear        & \mswRECAPNear        & \numRECAPNearMSW          \\
  \ Unrec.   & \unrecFOIA &  \unrecOIG         & \unrecDNSA          & \unrecRECAP          & \numRECAPunk              \\
  \ Exact Word     & 4,694             & 455              & \mswDNSAExact       & \mswRECAPExact       & \numRECAPmsw              \\
  Vulnerable       & 711               & 58               & 9                   & 0                    & \numRECAPmsw              \\
  \midrule
    No Matches          & \numFOIAUnmatched & \numOIGUnmatched & 1                  & --                   & N/A                        \\
  Uniq. Matches      & \numFOIAUniq      & \numOIGUniq      & 5                  & --                   & N/A                        \\
  Avg. Matches        & \numFOIAavg       & \numOIGavg       & 393                  & --                   & N/A                        \\
  Med. Matches        & \numFOIAmed       & \numOIGmed       & 1                  & --                   & N/A                        \\
  \bottomrule
\end{tabular}

\end{table}

\subsection{Results}
\label{sec:wildres}

\paragraph{Excising Redactions.}
Table~\ref{tab:wildres} reports our findings on the security of excising redactions.
The \emph{vulnerable} row reports the result of our redaction classification methodology.
We identified 711 vulnerable excising redactions in the FOIA corpus, 58 in the OIG corpus, 9 in the DNSA corpus, and none in rRECAP (mentioned above).
The lower half of the table reports statistics on the number of entries from our dictionary that were not determined to be impossible given observed leaked redaction information.

Due to our large combined dictionary size, we saw few unique matches (0.4\% of 769 FOIA and OIG redactions).
Deredaction still presents a serious threat, particularly in the case of constrained dictionaries.
For example, knowing a redaction is a U.S. congressperson's name would allow an attacker to deredact the name.

The size of the possible redacted texts after redaction is also skewed positively.
While FOIA's average case sees around a 3,000-fold reduction in the number of potential redacted texts, the majority of cases see at least a 14,000-fold reduction.

We find unmatched cases are common: false negatives arise if the word does not occur in the dictionary.
Manual analysis of nonexcising RECAP redactions found 28.2\% of names were of a form occurring in our dictionaries, leaving 71.7\% in other forms (e.g. first name, last name).
Our own false negative rate (54.7\%) is lower because, while we extensively tested our positioning scheme models, misleading results (false positives) appear to exist in approximately 17\% of cases.

\paragraph{Nonexcising Redactions.}
We ran \maxray's location algorithm for nonexcising redactions on all of RECAP and found \numRECAPNameredactions\ nonexcised redacted names in US court documents. 
The number of nonexcised redacted words was larger ($\approx$\num{1.4e5}).

\subsection{Validation}

We manually validated our results for excising redactions (where possible) by performing web searches for further information on the redacted document.
For example, in the case of an OIG investigation, we would perform a search related to the offense committed and organizations affiliated.
This process found our deredaction returned no false positives, though our false negative rate suggests some of our match sets may not include the redacted word.
False positives occur when \maxray\ returns a set of matches, but the redacted name is not in the dictionary.
Given the danger of false positives, deredaction should be understood as \emph{ruling out} possible candidate texts rather than determining the content of a redaction.

Nonexcising redactions provide ground truth, and we also used these redactions to validate our techniques.
For every nonexcised redacted name in RECAP present in our dictionary, we excised the name, i.e. we removed the redacted text but preserved non-redacted glyph shift information.
We then formed a set of matching names using \maxray, and we were successful in all cases, as the set of candidate texts returned by \maxray\ included the ground truth redacted word.

Two of these nonexcising redactions were of a name only (no other words on the line were redacted).
We report results for these redactions as \maxray's penultimate evaluation:

\emph{Mr. Hamilton.} 
The first redacted name had the form \emph{Mr. Hamilton} (actual name different). 
The result set contained 24 names and Mr. Hamilton's was the 14th most common based on US Census data. 
No matched name had the title \emph{Ms.}.

\emph{Ms. Schuyler.} 
The second redacted name had the form \emph{Ms. Schuyler} (actual name different). 
The result set contained 210 names and Ms. Schuyler's was the 127th most common based on US Census data. 
No matched name had the title \emph{Mr.}.

\section{Discussion}
\label{sec:disc}

In the prior sections, we demonstrated sub-pixel sized glyph position shifts, imperceptible to the human eye, can break text redactions.
We found no non-rasterizing redaction tools address these leaks and two of the tools, PDFescape Online and PDFzorro, do not even excise redacted glyphs.
There are at least 778 vulnerable excising redactions in FOIA, OIG, and DNSA PDF corpora and more than 700 publicly accessible court documents with nonexcising redactions.

We have provided a list of the nonexcising redactions and associated case numbers to RECAP and the US Court system.
We also notified other affected parties of the leaks and provided them with the tools to fix them.
In particular, our results show redacting a name from a PDF is not secure and new redaction practices should be adopted (Sec.~\ref{sec:rec-prac}).

\subsection{Defenses}
\label{sec:defenses}

Excising redactions may be safeguarded.
The official NSA guidelines for redaction of Microsoft Word generated PDFs are to change the content of the original Word document so that information regarding the sensitive name is destroyed (i.e. by changing the name to the letter ``x'')~\cite{nsaRedact}.
Below, we describe five alternative defenses against excising redaction vulnerabilities.
We are releasing parts of the \maxray\ tool suite that can be used to identify non-excising redactions, and tools to repair such redactions. 
We are \emph{not} releasing the Word document models that can be used to attack vulnerable excising redactions.

\paragraph{Glyph Shift Discretization.}
Shifting scheme noise may be added to a line without affecting visual fidelity.\footnote{
    Recall each text space unit is 1/6,000~in (0.0042~mm) (Sec.~\ref{sec:pdftext}).
} 
Alternatively, each shift could be rounded to a discrete interval, e.g. 0.1~mm.
If this noise is indistinguishable from legitimate glyph shift information, accounting for it would require an adversary to increase the set of accepted redacted text guesses.
Removing shifting scheme information altogether can also lower information leaks, though this is less visually appealing.

\paragraph{Document Layout and Redaction Obfuscation.}
Modifying the PDF commands used to render the box of the redaction complicates the process of automated redaction location.
For example, our nontrivial redaction location algorithm relies on identifying a black box between two US English words in order to avoid large numbers of false positives.

\paragraph{Increasing Redaction Width.} 
Redacting additional adjacent words can make deredaction more difficult, although this practice may not always compatible with legal mandates.
This defense works increases the number of words the attacker must guess. 
If the adjacent words are easy to infer, this is not a sufficient defense.

\paragraph{Adversarial Redactions.} 
One defense against deredaction is to change the document's text before it is redacted, for example, by replacing a sensitive name with the letter ``X''.
It is also possible to \emph{lie} to deredaction by changing the redacted content to something seemingly valid, potentially misinforming an adversary.

\paragraph{Rasterization.} 
Rasterization appears to be an effective defense against deredaction.
In many cases this defense is infeasible because it removes searchable text data from the document, however, performing OCR on the document post-redaction can act as a stopgap for this issue.
Rasterization algorithms may also modify or ignore certain glyph shifts,\footnote{We analyzed several rasterization tools, finding several had imperfect precision.} requiring the analyst to perform more reverse engineering to identify the specific rasterization tool used.

We performed an experiment to estimate the effects of rasterization on redaction security.
We chose ten name occurrences from court documents using the Word shifting scheme in 12 point Times New Roman.
For each selected text line, we substituted the name with each entry in our 235,560 name dictionary from Section~\ref{sec:wild}, redacted the entry, and calculated the redaction's width.
This resulted in 2,017 unique redaction widths, $\approx$11 bits of information leaked, on average.
Quantization to 300 DPI (20 text space units) resulted in 252 widths on average ($\approx$8 bits) and quantization to 600 DPI performed slightly worse with a result of 377 widths ($\approx$8.6 bits).

Unfortunately, this estimation is a lower bound.
In general, when a character is rasterized, the vector representation is converted to a ``mosaic'' of pixel values.
Whether a given pixel value is black, white, or in the case of anti-aliasing~\cite{romanyuk2015method}, some shade of gray, is dependent on how the graphical rendering specification is converted to a matrix of pixel values.
We leave an analysis of this precision loss due to particular rendering algorithms to future work.

\subsection{Recommended Practices}
\label{sec:rec-prac}

Redaction practices must account for concerns about \emph{document integrity}.
All the above measures modify the redacted document beyond simply removing text. 
In some contexts, particularly due to legal or regulatory reasons, this may not be acceptable. 
One of the main reasons for releasing redacted documents is to demonstrate \emph{transparency} while still protecting sensitive information.
Altering parts of the document outside the redaction alters this promise of authenticity.

It is technically possible to fix a non-excising redaction by removing the redacted text. 
The effect would be the same as if the document were redacted by an excising redaction tool. 
However, this also raises the issue of authenticity of done by a third party (e.g. document repository operator), because this necessarily means modifying the original document without the author's involvement.

In cases where integrity requirements may be relaxed, the NSA-recommended practice of altering the original document to replace the redacted text with meaningless text, e.g. REDACTED, provides the highest level of security. 

In cases where the underlying text may not be changed, we offer the following two suggestions.
First, we note that \textbf{redacting a name from a PDF is not secure}.
If a name occurs on a line of text, the entire line should be redacted, if possible, or care should be taken to ensure that enough of the surrounding words are redacted to make deredaction unlikely.
Second, if redacting more text is not possible, the width of the redaction should be quantized to a fixed value, and any glyph shifts should be removed.
While this may make the file less aesthetically pleasing, it is necessary for the security of redactions.

\subsection{Ethics}
\label{sec:ethics}

We carefully considered the balance between the benefits of public awareness and potential risks of misuse present in this research.
To the best of our abilities we are attempting to maximize the benefits to society and minimize the harm to individuals in the publication of this work.
In this spirit we have released open-source toolkits for protecting vulnerable redactions (Sec.~\ref{sec:maxray}).

All the documents studied are in the public domain.
So long as copies of these PDFs exist, they pose a risk to individuals' privacy.
During our evaluation, we ensured all deredaction was performed on an isolated and hardened server and that no identifying information exited this server.
We deleted the documents from our server and associated data after evaluation and notifying affected parties.

We have notified Microsoft and Adobe of our discoveries: although we understand they are \emph{not responsible} for the vulnerabilities, changes to their tools could help protect future PDF redactions.
We have also reached out to the PDF Association regarding the provision of guidance for redaction application implementers.

We reached out to the two redaction toolkits with full text redaction leaks and have been assigned CVE-2022-30350 and CVE-2022-30351.
We have also performed an extensive notification of the US Courts, the Free Law Project, the US Department of Justice, and the Office of Inspector General.
Our discussions with these groups are ongoing and include technical support, code, and free consulting regarding remedation and prevention efforts.

Careful deliberation has led us to determine that the present paper will motivate further remediation efforts rather than encourage attacks on redacted documents.
By bringing attention to the problem of redaction, this paper empowers affected individuals to advocate for the protection of their redacted information.

%
%
%

\section{Related Work}
\label{sec:related}

Our work is the first to consider the role the PDF glyph positioning information plays in redaction security.
We are also the first to present an algorithmic attack on redactions where the underlying text is removed and replaced with a black box.

\paragraph{Digital Redaction}
However, we are not the first work to discuss digital redaction.
Forrester and Irwin~\cite{forrester2005investigation} discuss trivial redactions and unscrubbed metadata such as the Producer field of PDF documents but do not mention glyph positioning based deredaction.
Hill et al., used hidden Markov models to recover text obscured either by mosaic pixelization or a related tactic, e.g. Gaussian Blur~\cite{hill2016effectiveness}.
While M{\"u}ller et al.~\cite{muller2021processing} do not explicitly tackle redaction, they discuss hidden information present in PDF documents, specifically PDF document revision information and author name metadata.
Beyond PDF document redaction, other file formats may also be deredacted: Murdoch and Dornseif~\cite{murdochmisc} discuss how cropped JPEGs can preserve uncropped image information.

\paragraph{Text Redaction Attacks}
The primary predecessor to our work is Lopresti and Spitz~\cite{lopresti2004quantifying}, which presents a manual technique for matching glyphs to a redaction's width in a raster image of text.
The authors attempted to use natural language processing to predict redacted words, something our work found imprecise given current models.
The Lopresti and Spitz work also conflates document glyph position specifications with TTF glyph widths and assumes both are equivalent to a raster document's character widths.
This presents two problems:

First, a rasterization workflow may change a document's glyph positioning and physical printing may not be a pixel-perfect reproduction of the digital document.
While the accurate representation of documents is the whole \emph{point} of PDF, a few of our tests found glyph positioning operations were not always honored.
Behavior and information from any \emph{singular} tool may or may or may not comply with the PDF standard, either because of software bugs or because the developers were too lazy to implement support for a specific glyph positioning operation.

Second, TTF glyph widths do not necessarily equate with PDF document or raster glyph widths.
TTF is only one of five types of fonts supported by PDF.
The Lopresti and Spitz techniques also rely on (potentially inaccurate) human determination of glyph positions.
Recall the present work provides a fully automatic deredaction method, a precise analysis of leaked information, and a clear measurement of this problem's prevalence in real documents.

\paragraph{Efforts by Government Organizations}
Outside the scientific community, some government agencies have studied redaction vulnerabilities.
The Australian Cyber Security Center~\cite{australiaRedact} analyzed Adobe Acrobat 2017's redaction security and considered several features including encryption, CMap leaks, redactions of text metadata, images, revision metadata, and form metadata.
However, this work does not address glyph positioning information leaks and incorrectly determines that Adobe Acrobat leaks no redacted information.
The National Security Agency's redaction guide~\cite{nsaRedact} does not mention glyph positioning information but notes any underlying redacted text should be removed from the document before producing a PDF.
Changing the underlying text before redaction is a great defense against deredaction attacks, however, we found redactors do not always follow this advisory.


\paragraph{High Profile Broken Redactions}
There have been many other cases of poorly redacted, high-profile documents.
Trivial redactions have been found in classified US Military documents~\cite{topSecretTrivial}, Manafort case documents~\cite{manafortTrivial}, and documents relating to Larry Page's house~\cite{larryPageTrivial}.
The AstraZeneca Contract with the EU disclosed funding information in the PDF's bookmarks~\cite{astraZenica}.
One Ghislaine Maxwell interview leaked redacted information on President Clinton in the document's index~\cite{epsteinTrivial}.


\section{Conclusion}
\label{sec:concl}

In the course of this work, we developed \maxray, a tool capable of locating and breaking redactions in thousands of PDF documents.
We found, for example, that redacting a surname from a PDF generated by Microsoft Word set using 10-point Calibri leaves enough residual information to uniquely identify the name in \puwavgpctnytlnwxxibx\ of all cases.
We surveyed the behavior of 11 different PDF redaction software tool-kits and found the majority do not defend against our attacks.
We discovered over 6,000 nonexcising redactions in US court documents and broke 348 excising redactions in Office of Inspector General reports and Freedom of Information Act requests.
This paper's findings are a lower bound on the extent of vulnerable redactions.


\bibliographystyle{plain}
\bibliography{local}

\begin{thebibliography}{10}

\bibitem{usCensus}
{Surnames Occuring 100 or More Times}.
\newblock
  \url{https://www.census.gov/topics/population/genealogy/data/2000_surnames.html},
  2000.

\bibitem{nsaRedact}
{Redacting with Confidence: How to Safely Publish Sanitized Reports Converted
  From Word to PDF}.
\newblock 2005.

\bibitem{australiaRedact}
{An Examination of the Redaction Functionality of Adobe Acrobat Pro DC 2017}.
\newblock
  \url{https://www.cyber.gov.au/acsc/view-all-content/publications/examination-redaction-functionality-adobe-acrobat-pro-dc-2017},
  2017.

\bibitem{oigReports}
{All Inspector General Reports in one Place}, 2021.

\bibitem{ssaNames}
{Beyond the Top 1,000 Names}.
\newblock \url{https://www.ssa.gov/oact/babynames/limits.html}, 2021.

\bibitem{demonymList}
{Demonym}.
\newblock \url{https://en.wikipedia.org/wiki/Demonym}, 2021.

\bibitem{dnsaSite}
{Digital National Security Archive}.
\newblock \url{https://nsarchive.gwu.edu/digital-national-security-archive},
  2021.

\bibitem{demonymCntry}
{List of Adjectival and Demonymic Forms for Countries and Nations}.
\newblock
  \url{https://en.wikipedia.org/wiki/List_of_adjectival_and_demonymic_forms_for_countries_and_nations},
  2021.

\bibitem{altCntry}
{List of Alternative Country Names}.
\newblock
  \url{https://en.wikipedia.org/wiki/List_of_alternative_country_names}, 2021.

\bibitem{geoNet}
{NGA GEOnet Names Server (GNS)}.
\newblock \url{https://geonames.nga.mil/gns/html/}, 2021.

\bibitem{pacerSite}
{Public Access to Court Electronic Records}.
\newblock \url{https://pacer.uscourts.gov/}, 2021.

\bibitem{govattic}
{The Government Attic}, 2021.

\bibitem{xrayTool}
{X-Ray Bad Redaction Detector}.
\newblock \url{https://free.law/projects/x-ray}, 2021.

\bibitem{libreOffice}
{LibreOffice Github}.
\newblock \url{https://github.com/LibreOffice}, 2022.

\bibitem{ncVoterData}
{North Carolina Voter Data}.
\newblock \url{https://www.ncsbe.gov/results-data/voter-registration-data},
  {2022}.

\bibitem{ohVoterData}
{Ohio Voter Data Download}.
\newblock
  \url{https://www6.ohiosos.gov/ords/f?p=VOTERFTP:STWD:::#stwdVtrFiles},
  {2022}.

\bibitem{latexSource}
{The TeX Live SVN}.
\newblock \url={https://www.tug.org/texlive/svn/}, 2022.

\bibitem{waVoterData}
{Washington Voter Registration Database Extract}.
\newblock \url{https://www.sos.wa.gov/elections/vrdb/extract-requests.aspx},
  {2022}.

\bibitem{pdfTwo}
PDF Association.
\newblock {\em {ISO 32000-2}}.
\newblock 2017.

\bibitem{manafortTrivial}
Natasha Bertrand.
\newblock {Manafort’s Own Lawyers May Have Hastened His Downfall}.
\newblock
  \url{https://www.theatlantic.com/politics/archive/2019/01/paul-manafort-lawyers-failed-to-redact-documents/579910/},
  2019.

\bibitem{cover-and-thomas}
Thomas~M. Cover and Joy~A. Thomas.
\newblock {\em {Elements of Information Theory}}.
\newblock 2 edition, 2006.

\bibitem{embarassingRedact}
Herbert Dixon.
\newblock {Embarrassing Redaction Failures}.
\newblock
  \url{https://www.americanbar.org/groups/judicial/publications/judges_journal/2019/spring/embarrassing-redaction-failures/},
  May 2019.

\bibitem{forrester2005investigation}
Jock Forrester and Barry Irwin.
\newblock {An Investigation into Unintentional Information Leakage through
  Electronic Publication}.
\newblock {\em Information Security South Africa}, 2005.

\bibitem{hill2016effectiveness}
Steven Hill, Zhimin Zhou, Lawrence Saul, and Hovav Shacham.
\newblock {On the (In) Effectiveness of Mosaicing and Blurring as Tools for
  Document Redaction}.
\newblock {\em Proceedings on Privacy Enhancing Technologies},
  2016(4):403--417, 2016.

\bibitem{timblee}
Timothy Lee.
\newblock {What Gets Redacted in Pacer?}
\newblock
  \url{https://freedom-to-tinker.com/2011/06/16/what-gets-redacted-pacer/},
  2011.

\bibitem{epsteinTrivial}
Josh Levin, Aaron Mak, and Jonathan Fischer.
\newblock {We Cracked the Redactions in the Ghislaine Maxwell Deposition}.
\newblock
  \url{https://slate.com/news-and-politics/2020/10/ghislaine-maxwell-deposition-redactions-epstein-how-to-crack.html},
  2020.

\bibitem{lopresti2004quantifying}
Daniel Lopresti and A~Lawrence Spitz.
\newblock {Quantifying Information Leakage in Document Redaction}.
\newblock In {\em Proceedings of the 1st ACM workshop on Hardcopy document
  processing}, pages 63--69, 2004.

\bibitem{larryPageTrivial}
Cade Metz.
\newblock {Privacy Watchdog Hoists Google by Its Own Petard}.
\newblock \url{https://www.theregister.com/2008/08/01/nlpc_outs_larry_page/},
  2008.

\bibitem{muller2021processing}
Jens M{\"u}ller, Dominik Noss, Christian Mainka, Vladislav Mladenov, and
  J{\"o}rg Schwenk.
\newblock {Processing Dangerous Paths}.
\newblock NDSS, 2021.

\bibitem{murdochmisc}
Steven Murdoch and Maximillian Dornseif.
\newblock {Far More Than You Ever Wanted To Tell: Hidden Data in Internet
  Published Documents}.
\newblock
  \url{http://irc.smurfnet.ch/events/CCC/congress/21c3/papers/271\%20Hidden\%20Data\%20in\%20Internet\%20Published\%20Documents.pdf},
  2004.

\bibitem{astraZenica}
Nikolaj Nielsen.
\newblock {EU Admits Redaction Error in AstraZeneca Contract}.
\newblock \url{https://euobserver.com/coronavirus/150799}, 2021.

\bibitem{romanyuk2015method}
Olexander~N Romanyuk, Sergii~V Pavlov, Olexander~V Melnyk, Sergii~O Romanyuk,
  Andrzej Smolarz, and Madina Bazarova.
\newblock {Method of Anti-Aliasing with the Use of the New Pixel Model}.
\newblock In {\em Optical Fibers and Their Applications 2015}, volume 9816,
  pages 274--278. SPIE, 2015.

\bibitem{nytCorp}
Evan Sandhaus.
\newblock {The New York Times Annotated Corpus}.
\newblock \url{https://catalog.ldc.upenn.edu/LDC2008T19}, 2008.

\bibitem{topSecretTrivial}
David Willey.
\newblock {Italy Media Reveals Iraq Details}.
\newblock \url{http://news.bbc.co.uk/1/hi/world/europe/4504589.stm}, 2005.

\end{thebibliography}


\appendix
\label{sec:appdx}

\section{Mutual Information Calculation.}
\label{appdx:mutinf}
The amount of information leaked by the redacted document $Y$ about the redacted word $X$ is given by the mutual information $I(X,Y)$.
Let $L$ be a random variable representing the location of the occurrence of the dictionary word chosen in our first evaluation step, and $H$ denote entropy.
Note that $H(Y|L,X) = 0$ (there is no uncertainty about $Y$ given $L$ and $X$), since the formatting and redaction in steps 3 and 4 are deterministic, and $H(L|X,Y) = H(L|Y) = 0$ (there is no uncertainty about the location of the redaction given the document after redaction). 
Using these two facts,

\begingroup\small
\begin{align*}
I(X&;Y) = \\
&= H(X) - H(X|Y) \\
&= H(X) - H(X,Y) + H(Y) \\
&= H(X) - H(L,X,Y) + H(L|X,Y) + H(L,Y) - H(L|Y) \\
&= H(X) - H(Y|L,X) - H(L,X) + 0 + H(L,Y) - 0 \\
&= H(X) - 0 - H(L) - H(X) + H(Y|L) + H(L) \\
&= H(Y|L) \\
&= \sum_\ell Pr[L=\ell]\cdot H(Y|L=\ell).
\end{align*}
\endgroup

Thus, $I(X;Y)$ is the average, taken over redaction locations, of the entropy of $Y$, that is, the entropy of the distribution of possible documents after redaction. 
For a large corpus and large dictionaries, calculating this quantity exactly is expensive. 
Instead of calculating the exact value, we sample $H(Y|L=\ell)$ for several initial word choices.

\section{Dictionary Construction}
\label{sec:dictdetail}

This paper used several dictionaries.
In order to ensure the quality of the guessed text, we manually sourced and refined the following:

\begin{enumerate}
\item \textit{Word} was sourced from the linux spellchecker: \url{/usr/share/dict/american-english} filtered for pronouns (first letter upper case), stop words sans pronouns, and words containing punctuation and digits. Single quotes were made to be curly possessive and straight to match the document under study.
\item \textit{FN} were sourced from from the US social security agency~\cite{ssaNames} starting from 1880. For both this and the last name dictionary names like ``McCarthy'' were made to be correctly capitalized.
\item \textit{LN} were sourced from the year 2000 US Census~\cite{usCensus}.
\item \textit{Ctry} was sourced from wikipedia lists of countries and alternative country names~\cite{altCntry,demonymCntry}, then refined by hand to ensure completeness with repect to potential forms, e.g. United States vs. The United States. For fun we added intialisms.
\item \textit{Rgn} is from the NGA GEOnet Names Server (GNS)~\cite{geoNet}. We filter for full name and BGN-approved local official name. We restrict our results to A, P,and L feature categories.
\item \textit{Natl} is sourced from wikipedia lists of demonyms~\cite{demonymCntry,demonymList}.
\end{enumerate}

\section{Leaked Information for 12 Point Fonts}\label{sec:12ptfonts-appdx}

\begin{table*}[h!]
    \caption{Leaked 12 point font information. 12 point font results for unadjusted and monospaced schemes are the same as Table~\ref{tab:nyt10}.}
    \label{tab:12ptfonts}
    \centering
    \begin{tabular}{l}
      \newcommand\nytxunknown{\raisebox{1.25pt}{---}}
\begin{tabular}{lrr@{\hspace{3pt}}rr@{\hspace{3pt}}rr@{\hspace{3pt}}rr@{\hspace{3pt}}rr@{\hspace{3pt}}rr@{\hspace{3pt}}rr}
\toprule
  & &
  \multicolumn{6}{c}{Leaked information (bits)}
  & \multicolumn{6}{c}{Probability unique match} \\
  \cmidrule(lr){3-8}\cmidrule(lr){9-14}
\multicolumn{2}{l}{\emph{\textbf{Distr}}} 
& \multicolumn{2}{c}{{\tfontname}}
& \multicolumn{2}{c}{{\afontname}}
& \multicolumn{2}{c}{{\bfontname}}
& \multicolumn{2}{c}{{\tfontname}}
& \multicolumn{2}{c}{{\afontname}}
& \multicolumn{2}{c}{{\bfontname}}
\\
& 
  \emph{Dict}
& {\wxiishortname} & {\wxxishortname}
& {\wxiishortname} & {\wxxishortname}
& {\wxiishortname} & {\wxxishortname}
& {\wxiishortname} & {\wxxishortname}
& {\wxiishortname} & {\wxxishortname}
& {\wxiishortname} & {\wxxishortname}
\\
\midrule
\multicolumn{6}{l}{\emph{\textbf{\uniname}}}
\\
& \emph{\acrnname}
  & 10.5 & 9.4 & 10.5 & 9.5 & 14.2 & 14.1 & <1\% & <1\% & <1\% & <1\% & <1\% & <1\%
\\
& \emph{\wordname}
  & 11.9 & 11.1 & 12.2 & 11.3 & 14.7 & 14.6 & 6\% & 3\% & 9\% & 5\% & 38\% & 37\%
\\
& \emph{\ctryname}
  & 9.0 & 9.0 & 8.9 & 8.9 & 9.1 & 9.1 & 88\% & 84\% & 83\% & 79\% & 95\% & 95\%
\\
& \emph{\rgnname}
  & 14.4 & 13.6 & 14.1 & 13.3 & 17.1 & 17.0 & 5\% & 3\% & 4\% & 3\% & 10\% & 9\%
\\
& \emph{\natlname}
  & 8.5 & 8.4 & 8.5 & 8.4 & 8.7 & 8.7 & 79\% & 74\% & 78\% & 72\% & 96\% & 96\%
\\
& \emph{\fnname}
  & 11.0 & 10.1 & 10.9 & 10.2 & 14.2 & 14.2 & 3\% & 1\% & 3\% & 2\% & 18\% & 17\%
\\
& \emph{\lnname}
  & 11.6 & 10.8 & 11.9 & 11.1 & 15.0 & 14.9 & 2\% & 1\% & 3\% & 2\% & 19\% & 18\%
\\
& \emph{\filnname}
 & 12.5 & 11.6 & 12.7 & 11.9 & 16.1 & 16.0 & 2\% & 1\% & 3\% & 2\% & 15\% & 14\%
\\
& \emph{\fixlnname}
  & 12.6 & 11.7 & 12.9 & 12.1 & 16.3 & 16.2 & <1\% & <1\% & <1\% & <1\% & 4\% & 3\%
\\
& \emph{\fnlnname}
 & 14.1 & 13.1 & 14.4 & 13.6 & 17.1 & 17.1 & 3\% & 2\% & 3\% & 2\% & 10\% & 10\%
\\
& \emph{\fnxlnname}
  & 12.7 & 11.8 & 12.8 & 12.1 & 16.9 & 16.9 & <1\% & <1\% & <1\% & <1\% & <1\% & <1\%
\\
\multicolumn{6}{l}{\emph{\textbf{Text frequency distr.}}}
\\
& \emph{\acrnname}
  & 8.2 & 7.8 & 8.1 & 7.7 & 9.6 & 9.6 & 3\% & 3\% & 3\% & 2\% & 9\% & 9\%
\\
& \emph{\wordname}
  & 10.3 & 9.8 & 10.5 & 9.8 & 11.9 & 11.9 & 4\% & 1\% & 5\% & 2\% & 38\% & 36\%
\\
& \emph{\ctryname}
& 5.8 & 5.7 & 5.8 & 5.7 & 5.8 & 5.8 & 83\% & 79\% & 82\% & 73\% & 91\% & 92\%
\\
& \emph{\rgnname}
& 9.3 & 9.0 & 9.1 & 8.9 & 10.0 & 9.9 & 1\% & <1\% & 1\% & <1\% & 5\% & 5\%
\\
& \emph{\natlname}
& 5.3 & 5.3 & 5.3 & 5.3 & 5.4 & 5.4 & 78\% & 75\% & 78\% & 71\% & 97\% & 98\%
\\
& \emph{\fnname}
 & 9.0 & 8.6 & 8.8 & 8.5 & 10.0 & 10.0 & 74\% & 67\% & 71\% & 66\% & 93\% & 93\%
\\
& \emph{\lnname}
 & 10.4 & 9.8 & 10.5 & 9.9 & 12.3 & 12.3 & 54\% & 47\% & 55\% & 48\% & 83\% & 83\%
\\
& \emph{\filnname}
 & 11.9 & 11.1 & 11.9 & 11.2 & 14.7 & 14.7 & 25\% & 18\% & 23\% & 19\% & 58\% & 57\%
\\
& \emph{\fnlnname}
 & 13.6 & 12.7 & 13.8 & 13.1 & 16.2 & 16.2 & 20\% & 15\% & 20\% & 17\% & 41\% & 41\%
\\
\bottomrule
\end{tabular}

    \end{tabular}
\end{table*}

Overall, we found 12 point font sizes leak only slightly less information in dependent schemes than 10 point font sizes.
However, the difference is not significant: we report our results in Table~\ref{tab:12ptfonts}.
For Calibri, the larger font size leaked \emph{more} information, because the larger font size emphasizes the small errors accounted for by Word's glyph shifting scheme.

Note that we do not reproduce the independent scheme results for this table since they are identical to Table~\ref{tab:nyt10}.

\section{Redaction Location}\label{sec:redact-loc-appdx}

\begin{table}[h!]
  \centering
  \caption{Accuracy of locating excising redactions using a box-walk routine vs. Timothy B. Lee's method, which records graphics state draw commands and looks for rectangles. Here we seperate redactions into ``easy'' and ``hard'' categories, where ``hard'' redactions are, for example, boxes drawn in the document with no surrounding text, or redactions that extend across and entire line and thus are too long to attack.}
  \label{tab:redactfind}
  
  \begin{tabular}{l@{\hskip -1cm}lrrrr}
    \toprule
 &                       & \emph{DNSA} & \emph{FOIA} & \emph{Govt.} & \emph{rRECAP} \\
    \midrule
    \ \textbf{Box}                                                                  \\
 & Easy Identified       &          9   &         84  &          61  &         0    \\
 & False Negative Easy   &          0   &         3   &          0   &         0    \\
 & Hard Identified       &          9   &         34  &          71  &         1    \\
 & False Positives       &          0   &         0   &          0   &         0    \\
    \ \textbf{Lee}                                                                 \\
 & Easy Identified       &          1   &         2   &          25  &         0    \\
 & False Negative Easy   &          0   &         0   &          0   &         0    \\
 & Hard Identified       &          10  &         53  &          33  &         1    \\
 & False Positives       &          0   &         6   &          26  &         57   \\
    \bottomrule
  \end{tabular}

\end{table}

In Table~\ref{tab:redactfind} we report the comparison between our algorithm for locating excising redactions and the one provided by Timothy B. Lee~\cite{timblee}.
We give an outline of this algorithm in Figure~\ref{fig:nontriv-alg}.
This algorithm, we note, is optimized to find boxes with respect to \emph{other non-redacted words} on the page, whereas Lee's method looks for rectangle draw commands.
Lee's method is therefore better at detecting redactions with no surrounding text, but these redactions are less useful for the deredaction attacks on excising redactions presented in this paper.
We also added a second step to the x-ray algorithm~\cite{xrayTool} for locating nonexcising redactions which removed a large number of false positives from the results (algorithm in Figure~\ref{fig:xray-alg}).

These bash scripts call into C, python, and ruby for various subroutines.
We do not use open source tools other than Poppler for this work, because in general we found most tools were inaccurate.
We performed extensive validation (mentioned throughout this paper) to ensure our recovery of PDF glyph positioning information was precise and correct.

The ``pts'' command is our own, and handles lifting a PDF specification into an intermediate representation consisting of glyph coordinates at the scale of text space units, mentioned briefly in Section~\ref{sec:pdftext}.
\url{find\_redaction\_box.py} and \url{trivial\_redaction\_loc.py} work as described earlier in this paper.
The former checks every sufficiently large space between two words for a redaction rectangle and the latter compares the pixels of the two rasterized images at glyph coordinates piped in by \url{get\_word\_coords.rb}.

\lstset{ %
language=bash,                
basicstyle=\ttfamily\footnotesize,       
numbers=none,                   
numberstyle=\ttfamily\footnotesize,      
stepnumber=1,                   
numbersep=5pt,                  
backgroundcolor=\color{white},  
showspaces=false,               
showstringspaces=false,         
showtabs=false,                 
frame=single,           
tabsize=2,          
captionpos=b,           
breaklines=true,        
breakatwhitespace=false,    
escapeinside={\%*}{*)}          
}

\begin{figure}[h!]
\begin{lstlisting}
tf=$(mktemp -u)
# Removes images from PDF
cpdf -draft "$1" -o "$tf".pdf-0
$SRC/painting/remove-tjs.sh "$tf".pdf-0 "$tf".pdf
convert -background 'rgb(255,255,255)' \
        -colorspace gray -normalize \
        -density 300 -quality 100 -depth 8 \
        "$tf".pdf "$tf".ppm 2>/dev/null
# parses out all the PDF text state and
# walks redaction boxes
$SRC/c-src/pts "$1" 1 |
    $SRC/location/find_redaction_box.py "$tf".ppm
\end{lstlisting}
\caption{Nontrivial redaction location algorithm}
\label{fig:nontriv-alg}
\end{figure}

\begin{figure}[h!]
\begin{lstlisting}
# convert all text to black
"$DIR"/lib/camlpdf -blacktext "$1" -o "$1"-a
# remove all text from one pdf
gs -q -o "$1"-b \
   -sDEVICE=pdfwrite -dFILTERTEXT "$1"-a
# create two ppms
pdftoppm -singlefile -r 300 "$1"-a "$1"-a
pdftoppm -singlefile -r 300 "$1"-b "$1"-b
# get coordinates of each word and compare pixels for differences
"$DIR"/get_word_coords.rb "$1"-a |
  "$DIR"/trivial_redaction_loc.py \
  "$1"-a.ppm "$1"-b.ppm
\end{lstlisting}
\caption{Two-pass algorithm for locating trivial redactions}
\label{fig:xray-alg}
\end{figure}

\section{PDF Workflows}
\label{appdx:flows}

We performed an analysis of several PDF document production workflows across a variety of operating systems and software.
Our results are reported in Table~\ref{tab:pdfflows}.
This table demonstrates the wide variety of glyph shifting schemes in different software sets.
Each row grouping represents a different general environment and software creator for the production of the PDF document, and each row represents a specific workflow---mostly which buttons are pressed during the PDF's creation.

\begin{table*}
  \centering
  \caption{Several possible PDF workflows. Each entry in the column on the left should be read from left to right, indicating the stages used to produce the document. For example, ``Edge/Firefox, Print/PDFViewer, Save'' can be interpreted as ``using the edge or firefox browsers’ PDF viewer or print dialog, hit save'' and results in the right column's displacements.}
  \label{tab:pdfflows}
    \resizebox{\columnwidth}{!}{\begin{tabular}{l@{\hskip -4cm}r}
  \toprule
  Software Stack, Workflow & Description of Positioning Scheme \\
  \midrule \multicolumn{2}{l}{\emph{\textbf{Word 365, Windows 10}}} \\
      {Edge/Firefox, Save} &
      [(E)7(xhi)7(bi)7(t)7( A)-6(. )] \\
      {Edge/Firefox, Edited, Save} &
      [(E)7(xhi)7(bi)7(t)7( )-1.75(A. )] \\
      {Edge/Firefox, Print/PDFViewer, Save} &
      [(E)-13(xhi)28(bi)28(t)-34( A)-27(. )] \\
      {Edge/Firefox, Edited, Print/PDFViewer, Save} &
      [(E)-13(xhi)28(bi)28(t)-34( )-3.17(A. )] \\
      {Edge, Edited, Print/PDFViewer, Print} &
      Stream of Vector Graphics Commands \\
      {Firefox, Print, Save} &
      Stream of Vector Graphics Commands \\
      {Firefox, Print, Print} &
      [(Exhi)7(bi)7(t)7( A)-8(.)-20( )] \\
      {Chrome} &
      Subset of Edge Outputs \\
      {Desktop, Save (all flows, including loaded again in Adobe)} &
      [(Exhibi)-2(t A. )] \\
      {Desktop, Edited Save (all flows)} &
      [(Exhibi)-2(t )-2.67(A.)-2( )] \\
      {Desktop, Edited/Unedited Print} &
      [(E)-0.8398(xhi)-0.8320(bi)-0.8320(t)-0.8320( )-10(A)-0.1680(.)10( )] \\
  \midrule \multicolumn{2}{l}{\emph{\textbf{Word 365, Mac OS}}} \\
      {Safari, Chrome, Firefox, Edge} &
      Subset of Windows 10, Edge \\
      {Desktop, Save (all flows)} &
      [(E)0.2(xhi)0.2(bi)0.2(t)0.2( A)-0.2(.)-0.104( )] \\
      {Desktop, Edited Save (all flows)} &
      [(E)0.2(xhi)0.2(bi)0.2(t)0.2( )-0.136(A.)0.232( )] \\
  \midrule \multicolumn{2}{l}{\emph{\textbf{Word 2016, Windows 10}}} \\
      {Save} &
      [(Ex)-8(hibi)6(t A. )] \\
      {Save, Edited} &
      [(Ex)-8(hibi)6(t )-2.67(A.)-2( )] \\
      {Print} &
      [(E)-0.8398(x)-10(hi)-0.8320(bi)7(t)-0.8320( )-10(A)-0.1680(.)10( )] \\
  \midrule \multicolumn{2}{l}{\emph{\textbf{Google Docs, Windows 10}}} \\
      {Save (All flows)} &
      [(Exhibit A.)] \\
      {Save, Adobe, Save} &
      [(Exhibit )55.838(A. )], [(AAA)128.417(VVV)] \\
      {Chrome/Edge, Print, Adobe PDF Save} &
      [(0123435ÿÿÿ78ÿÿÿ)] (Actual embedded text is clobbered.) \\
      {Edge, Print, Print} &
      Stream of Vector Graphics Commands \\
      {Firefox, Download/Print, Firefox Viewer, Print, Save} &
      File rasterized \\
      {Firefox, Download/Print, Firefox Viewer, Print, Print} &
      Stream of Vector Graphics Commands \\
      {Firefox, Download/Print, Firefox Viewer, Print, Adobe} &
      Stream of Vector Graphics Commands (Adobe specific) \\
  \midrule \multicolumn{2}{l}{\emph{\textbf{Google Docs, Ubuntu 21.04}}} \\
      {Firefox/Chrome, Save (all flows)} &
      [(Exhibit A.)] \\
      {Firefox, Download, Firefox Viewer, Print} &
      File rasterized \\
      {Chrome, Print/System Print} &
      [(Exhibit\ \ \ A.\ \ \ )] (Spaces tripled) \\
  \midrule \multicolumn{2}{l}{\emph{\textbf{Google Docs, Mac OS}}} \\
      {Chrome/Firefox/Safari, Download/Print, Save} &
      [(Exhibit A. )] \\
      {Chrome, Print, Save as Adobe PDF} &
      Same as Ubuntu, Chrome, System Print \\
      {Chrome, Print, Preview/Sysdiag Print, Export} &
      [(E)0.2(hi)0.2(bi)0.2(\ \ \ A.\ \ \ )], (Spaces Tripled) \\ 
      {Firefox, Print, Preview, Export} &
      [(E)0.2(hi)0.2(bi)0.2( A. )] \\
      {Firefox, Download, Firefox Viewer, Print} &
      File rasterized \\
  \midrule \multicolumn{2}{l}{\emph{\textbf{Quartz PDFContext (Apple Pages), Mac OS}}} \\
      {Export/Print/Preview, Save} &
      [(E)0.2(xhi)0.2(bi)0.2(t)0.2( )55.2(A)-0.2(. )], [(AAA)129.0(VVV)] \\
      {Print, Save-as-Adobe-PDF} &
      [(Exhibit A. )], [(AAA)128.8(VVV)] \\
  \midrule {\emph{\textbf{Word 2012, Windows 10}}} &
  Same as Word 2016, Windows 10 \\
  \midrule {\emph{\textbf{All Word Versions, Windows 8.1 and Older}}} &
  Subset of Windows 10 \\
  \midrule {\emph{\textbf{No Disp. 600 DPI, Adobe OCR}}}  \\
    {OCR Only, Save} &
    [(E)-2.3(x)-2.3(h)-2.3(i)-2.3(b)-2.3(i)-2.3(t)-2.3( )\emph{[79.75 Td - 2.3 Tc]}(A)3.3(.)3.3( )3.3()] \\
    {Edit or OCR then Edit, Save} &
    [(E)-0.12(x)-0.12(h)-0.12(i)-0.12(b)-0.12(i)-0.12(t)-0.12( )\emph{[71.6 Td - 0.12 Tc]}(A. )] \\
  \bottomrule
\end{tabular}
}
\end{table*}

\section{The Acrobat Pro Glyph Shifting Scheme}
\label{sec:acro-pro-deets}

We can divide the effects of Acrobat Pro on PDF glyph shifts along into two functionalities: document editing and document OCR.
Clicking on the text of an existing PDF in Acrobat does not change any positioning information.
However, when a user edits a text object (typically one line of text) Acrobat sets all the object's glyph shifts to 0, making the scheme unadjusted.
During OCR, Acrobat (1) creates text objects and (2) adds glyph shifts to otherwise unadjusted text.
Acrobat's OCR algorithm creates glyph shift values by adding character and word spacing operators to each word, starting at the first letter of each word.
The former, \texttt{Tc}, adds a small shift to each character in the word.
The latter, \texttt{Td}, adjusts the word's x,y coordinates relative to the end of the prior word.

Redaction tools may remove these text positioning operators, making it impossible to infer the precise width of the redacted word.
This is because the redacted text's width (as given by unredacted glyphs) may not necessarily be equivalent to the total glyph shift used to represent the redaction.
However, we found these operators' parameters can be inferred via two sidechannels:

\begin{enumerate}
    \item \textbf{Tc}: This operator also applies to the successor space character of the word. 
        Since this space is rarely redacted, the applied Tc command remains in the PDF after redaction.
        This allows us to infer the precise spacing applied to the redacted text's characters, and therefore precisely infer the width of redacted text when used in combination with the Td operator sidechannel (discussed next).
        We validated this for all redaction tools in Section~\ref{sec:tools}.
    \item \textbf{Td}: All of Section~\ref{sec:tools}'s redaction tools draw a box after removing the glyphs for the redacted text.
        This black or white box's coordinates match those of the first glyph in the redacted word \emph{after} the Td adjustment, revealing the redaction's exact width.
        If redaction removes the word's predecessor space character, this sidechannel is removed.
        There may not exist a space character before the redacted information, however, in this case other information such as the left justification margin may be inferred from other text on the page and used instead.
\end{enumerate}

As a result, if a document is run through Acrobat's OCR before redaction and these sidechannels are not removed.
Redactions of documents produced by Acrobat's OCR are therefore not much more secure than PDFs with an unadjusted shifting scheme.


\lstset{ %
language=C++,                
basicstyle=\ttfamily\footnotesize,       
numbers=none,                   
numberstyle=\footnotesize,      
stepnumber=1,                   
numbersep=5pt,                  
backgroundcolor=\color{white},  
showspaces=false,               
showstringspaces=false,         
showtabs=false,                 
frame=single,           
tabsize=2,          
captionpos=b,           
breaklines=true,        
breakatwhitespace=false,    
escapeinside={\%*}{*)}          
}

\begin{figure}
\begin{lstlisting}
int i = 0;
int trackingAdj = 0;
do {
  int accumulatedDiff = 0;
  int lastNewAdj = 0;
  int totalWidth = 0;
  int amountAdjustedSoFar = 0;
  totalWidth += fontScaledWidths[i];
  int newAdjustment = PIXEL_W(totalWidth);
  accumulatedDiff = pixelWidths[i] - newAdjustment;
  pixelWidths[i] = amountAdjustedSoFar = lastNewAdj = newAdjustment;
  uncorrectedPixelWidths[i] = newAdjustment;
  i++;
  if (i == numChars)
    break;
  do {
    totalWidth += fontScaledWidths[i];
    if (totalWidth > WIDTH_BRKPOINT)
      break;
    int origAdjustment = pixelWidths[i];
    int newAdj = PIXEL_W(totalWidth);
    newAdj -= amountAdjustedSoFar;
    int adjustmentDifference = origAdjustment - newAdj;
    trackingAdj = adjustmentDifference - accumulatedDiff;
    if (adjustmentDifference != accumulatedDiff) {
      int v28 = trackingAdj & 1;
      int v50 = trackingAdj & 1;
      if (trackingAdj <= 0) {
        trackingAdj >>= 1;
        if (adjustmentDifference < -accumulatedDiff)
          trackingAdj += v50;
        if (-newAdj >= trackingAdj)
          trackingAdj = -newAdj;
      } else {
        trackingAdj >>= 1;
        if (accumulatedDiff < -adjustmentDifference)
          trackingAdj += v28;
        if (lastNewAdj < trackingAdj)
          trackingAdj = lastNewAdj;
      }
    }
    pixelWidths[i - 1] -= trackingAdj;
    int newTrackAdj = newAdj + trackingAdj;
    amountAdjustedSoFar = amountAdjustedSoFar + newAdj;
    accumulatedDiff = origAdjustment - (newTrackAdj);
    pixelWidths[i] = newTrackAdj;
    uncorrectedPixelWidths[i] = newTrackAdj;
    lastNewAdj = newAdj;
    i++;
  } while (i < numChars);
} while (i < numChars);
\end{lstlisting}
\caption{Word WYSIWYG width adjustment method.}
\label{fig:wysiwyg-alg}
\end{figure}

\begin{figure}
\begin{lstlisting}
if (msword_year <= 2016) {
  int leadingSpace = 1;
  float dev = 0;
  float ttf = 0;
  for (int j = i + 1; j < vs->size(); j++) {

    if (vs->getChar(j) == ' ' && leadingSpace) {
      continue;
    } else if (leadingSpace) {
      leadingSpace = 0;
    }

    float t = textSpaceWidths2007[j] / 1000;
    double d = deviceWidths[j] / msFSize2007;

    ttf += t;
    dev += d;
    double disp = ttf - dev;
    if ((disp > 0.003 || disp < -0.003) && i != vs->size() - 1) {
      int adj = disp * 1000 + 0.5;
      vs->setShift(j, adj);
      ttf = dev = 0;
    } else {
      vs->setShift(j, 0);
    }
  }
} else {
  int leadingSpace = 1;
  float dev = 0;
  float ttf = 0;
  float t = 0;
  float d = 0;
  double disp = 0;
  for (int j = i + 1; j < vs->size(); j++) {

    if (vs->getChar(j) == ' ' && leadingSpace) {
      continue;
    } else if (leadingSpace) {
      leadingSpace = 0;
    }

    t = textSpaceWidths2019[j] / 1000;
    d = deviceWidths[j] / msFSize2019;

    ttf += t;
    dev += d;
    disp = ttf - dev;

    if (((disp > 0.003) || (disp < -0.003)) && i != vs->size() - 1) {
      int adj = disp * 1000 + 0.5;
      vs->setShift(j, adj);
      ttf = dev = 0;
    } else {
      vs->setShift(j, 0);
    }
  }
}
\end{lstlisting}
\caption{Adjustment routine for Word's displacment scheme}
\label{fig:adjust-alg}
\end{figure}

\begin{figure}
\begin{lstlisting}
   std::vector<double> Office::computeEditAdjustments(VectorString *vs) {
       std::vector<double> editAdjustments;
       int totalDots = 0;
       double totalDevWidth = 0;
       for (int i = 0; i < numChars; i++) {
           totalDots += (i == numChars - 1) ? uncorrectedPixelWidths[i] : pixelWidths[i];
           totalDevWidth += (textSpaceWidths2019[i] - vs->getShift(i)) * (msFSize2019 / 1000);
           double realWidth, editedWidth, adj;
           realWidth = totalDevWidth + dotsToPdfUnits2019(600);
           editedWidth = roundToDigits(dotsToPdfUnits2019(totalDots + 600), 5);
           adj = (realWidth - editedWidth) / (msFSize2019 / 1000);
           editAdjustments.push_back(adj);
       }
       return editAdjustments;
   }
   std::vector<double> Office::edit(VectorString *vs, std::vector<double> eAdjs, int i) {
       double disp;
       disp = roundf(eAdjs[i] * (msFSize2019 / 1000) * 1000) / 1000 / (msFSize2019 / 1000);
       vs->addShift(i, disp);
       adjust(vs, i);
       return computeEditAdjustments(vs);
   }
   void Office::editSuffix(std::vector<double> savedEAdjs, VectorString savedVs, VectorString *suf) {
       double guessShift;
       double checkShift;
       double adjShift;
       bool matches = true;
       for (int i = 0;i < suf->size()-1;i++){
           int ind = savedVs.size() - suf->size() + i;
           guessShift = roundf(10 * savedVs.getShift(ind));
           checkShift = roundf(10 * suf->getShift(i));
           if (ind == savedVs.size() - 1) {
               savedVs.setShift(ind, suf->getShift(i));
               break;
           }
           adjShift = roundf(10*savedEAdjs[ind]);
           if (guessShift + adjShift == checkShift) {
               savedEAdjs = edit(&savedVs, savedEAdjs, ind);
           } }
       for (int i = 0;i < savedVs.size()-1;i++){
           if (savedVs.getChar(i) == L'-') {
               savedEAdjs = edit(&savedVs, savedEAdjs, i - 1);
               savedEAdjs = edit(&savedVs, savedEAdjs, i);
           } } }
\end{lstlisting}
\caption{Microsoft Word edit history shifting scheme algorithm.}
\label{fig:edit-alg}
\end{figure}

\begin{figure}
\begin{lstlisting}
for (int i = 0; i < numChars; i++) {
  fontScaledWidth = widths[i] * fontSize * 2;
  fontScaledWidths.push_back(fontScaledWidth);
  if (msword_year == 2007) {
    float textSpaceW =
        roundf(roundf(((float) widths[i]) / UNIT_TEXT_SPACE_2007 * 10000) / 10);
    textSpaceWidths2007.push_back(textSpaceW);
  } else {
    double textSpaceW = roundf(((double) widths[i]) / UNIT_TEXT_SPACE_2019 * 1000.0);
    textSpaceWidths2019.push_back(textSpaceW);
  }
  uncorrectedPixelWidths.push_back(0);
}
if (justif) {
  justify();
}
computeWYSIWYG();
\end{lstlisting}
\caption{Word initialization of widths.}
\label{fig:initalg}
\end{figure}

\section{Microsoft Word Shifting Scheme}
\label{sec:mic-word-deets}
While other shifting schemes leak proportional width information, any accumulation of information conditioned on redacted glyphs potentially leaks information.
Additional redacted information leaks occur because Microsoft Word's \emph{Save As PDF} tool converts between two glyph coordinate representations.
When users edit or create a Word document, they operate on a virtual document representation, internal to Word.
This representation determines how Word displays the document in the graphical user interface (GUI).
Note, however, we found the resolution of the user's screen and other environmental factors do not affect this virtual representation \emph{insofar as} the virtual representation is used for PDF generation.
Included in the virtual document's representation are a set of \emph{internal widths} used to represent glyphs' sizes.

When Word writes a PDF file, it translates the coordinates for glyphs in this virtual representation into their respective coordinates in the resultant PDF document.
The PDF file contains a separate glyph width mapping, typically matching the embedded TrueType font file.
A small amount of error exists between the two representations' rendering width specifications for glyphs.

A glyph's virtual width representation can be smaller or larger than that same glyph's PDF width specification.
To account for this error, Word's PDF document writer accumulates each glyph's width error from left to right across each line of text twice.
The first pass modifies the internal representation of each character's width according to the difference between the current internal width and the TTF width\footnote{
    This choice of TTF is Microsoft's choice. It here refers to the TTF file embedded in the PDF after the ``Save As PDF'' command.
    } of the line's prefix up to the current position, at around a 600 DPI resolution.
Note this is \emph{not to Print to PDF} workflow, so no part of this operation is dependent on the operating system's printer driver.

In the second pass, a second left-to-right accumulation occurs, accounting for error between the widths output by the first pass ad the TTF line width, converted to text space units (typicaly 1/1000th of the font size at 72 DPI).
When the accumulated error hits a three text space unit threshold, Word writes a shift to the PDF and resets the accumulator.
In summary, the font metrics of the line to the left of any given threshold value effect the accumulated error, leaking redacted information via the redaction's width and non-redacted glyph shifts.\footnote{
    Monospace fonts do not contain conversion width errors.
}
For example, a redacted glyph with a 10 text space unit error will overflow the accumulator and the glyph's effective PDF width will be its original advance width plus a shift between 7 and 13 units depending on the accumulator's state before overflow.

We note the line's specific character ordering affects the accumulator state as different orderings will lead to different accumulator overflow positions.
For example, the accumulated error could fluctuate between postive and negative three text space units for several characters and then overflow.
Both the point of overflow and the magnitude are conditioned on prior glyphs' modification of the accumulated error value.
For example, in Times New Roman, the digits 0 through 9 have no error and do not contribute to the accumulator.
However \emph{a lack} of a shift value can also leak information about the redacted content.
Additionally, due in part to the algorithm's first pass, smaller fonts leak more information as they have more characters per line and more threshold opportunities.

We validated \maxray's models for Word versions between 2007 and 2019 using several fonts on hundreds of lines of text from Wikipedia, hundreds of manual tests, and thousands of real-world cases in Section~\ref{sec:wild}.
Section~\ref{sec:wild} describes \maxray's process for correctly identifying Word shifting schemes.
We also note that Microsoft Word's shifting scheme metadata depends on the Word document's edit history.
Appendix Figure~\ref{fig:edit-alg} gives code for edit history modeling.

\subsubsection{Word Glyph Positioning Algorithms}\label{sec:word-disp-appdx}

The first routine (Figure~\ref{fig:initalg}) intializes a set of widths from two files, widths corresponds to TTF widths, and pixelWs are Word's internal widths for each character. It then initializes the WYSIWYG sizes.
The next routine (Figure~\ref{fig:wysiwyg-alg}) handles calculating the WYSIWYG widths for each character by accumulating errors between the two width representations, consisting of a tested loop that checks sets of runs of characters:
Then (Figure~\ref{fig:adjust-alg}), once an actual line of text needs to be adjusted, the deviceWidths initialized during the WYSIWYG modification are checked for overflow. 
The types of floating point variables, e.g. float vs. double, are precisely chosen and must be identical to the original algorithm or the result will be incorrect.

\section{Latex, LibreOffice, Other PDF Producers}
\label{sec:other-prod}

\textbf{LibreOffice} LibreOffice~\cite{libreOffice} includes a PDF writer as part of its Visual Class Library (VCL), and, at the time of writing, embeds shifts in the drawHorizontalGlyphs method of the PDFWriterImpl class. 
This method appears to be similar to Microsoft Word in function: shifts are applied whenever the native advance width of the PDF glyph does not match the Pixel x,y coordinates that LibreOffice uses internally, derived from the SalLayout (layout engine) class.
However, LibreOffice does not use a second width map for the characters in a given font, and therefore likely leaks less redacted information than Microsoft Office's shifting scheme.

\textbf{\LaTeX\ (TeX Live)}
The determination of shifts inside \LaTeX\ is dependent upon the particular flow that is used~\cite{latexSource}.
For the pdfTeX flow, interglyph shifts are decided by the pdf\_begin\_string procedure. 
The \LaTeX\ glyph placement algorithm is quite complex, however, our analysis did find one section of the code for determining shifts appears to round the difference between the current horizontal glyph render position and the start of the current text object.
Depending on the precision of the rounding, this could leak additional information about the order of glyphs used.
Other flows, such as XeLaTeX, can optionally include more complex calculations that determine how to shrink and expand glyphs to fit a certain amount of space and may leak large amounts of redacted information.
However, \LaTeX\ redactions are uncommon.

\textbf{Other Producers}
We found other producers, such as iText, where the default is unadjusted text objects, are accounted for by the schemes presented in the main body of this paper (Section~\ref{sec:pdftext}).

\subsection{Document Scanners}
\label{sec:doc-scanners}
Many document scanners also come with OCR features.
We examined the output of three different document scanners: a Canon MP Navigator EX, Xerox AltaLink C8145, and HP MFP M227fdn.
Of the three, the Canon and Xerox machines provided built-in software for document OCR.
We examined the Canon's glyph shifting scheme first, and found redactions on PDFs produced by this workflow to be equally as vulnerable to deredaction as unadjusted schemes.
The Xerox machine was not as simple: the Xerox machine's OCR text objects are positioned stochastically without space characters in between them.
After testing several redaction tools on the Xerox document, we found it was as vulnerable as a document with an unadjusted scheme as long as redaction was performed by removing glyphs rather than entire text objects.

\end{document}